\documentclass[5p,times,twocolumn]{elsarticle}

\usepackage{lineno,hyperref}
\usepackage{amsmath,amssymb,amsfonts}
\usepackage{textcomp}
\usepackage{bm}
\usepackage{array}
\usepackage{booktabs}
\usepackage{soul}
\usepackage{changes}
\definechangesauthor[name={Per cusse}, color=orange]{per}

\journal{Journal of Elsevier}

\begin{document}

\begin{frontmatter}

\title{Multi-level colonoscopy malignant tissue detection \\ with adversarial CAC-UNet}


\author[mymainaddress]{Chuang Zhu}
\ead{czhu@bupt.edu.cn}
\author[mymainaddress]{Ke Mei}
\author[mymainaddress]{Ting Peng}
\author[mymainaddress]{Yihao Luo}
\author[mymainaddress]{Jun Liu}
\author[mysecondaryaddress]{Ying Wang}
\author[mysecondaryaddress]{Mulan Jin}



\address[mymainaddress]{School of Information and Communication Engineering, Beijing University of Posts and Telecommunications, Beijing, China}
\address[mysecondaryaddress]{No. 8 Gongti South Road, Chaoyang District, Beijing, China}

\begin{abstract}
The automatic and objective medical diagnostic model can be valuable to achieve early cancer detection, and thus reducing the mortality rate. In this paper, we propose a highly efficient multi-level malignant tissue detection through the designed adversarial CAC-UNet. A patch-level model with a pre-prediction strategy and a malignancy area guided label smoothing is adopted to remove the negative WSIs, with which to lower the risk of false positive detection. For the selected key patches by multi-model ensemble, an adversarial context-aware and appearance consistency UNet (CAC-UNet) is designed to achieve robust segmentation. In CAC-UNet, mirror designed discriminators are able to seamlessly fuse the whole feature maps of the skillfully designed powerful backbone network without any information loss. Besides, a mask prior is further added to guide the accurate segmentation mask prediction through an extra mask-domain discriminator. The proposed scheme achieves the best results in \textit{MICCAI} DigestPath2019 challenge\footnote{\url{https://digestpath2019.grand-challenge.org/}} on colonoscopy tissue segmentation and classification task. The full implementation details and the trained models are available at \url{https://github.com/Raykoooo/CAC-UNet}.
\end{abstract}


\end{frontmatter}


\section{Introduction}
Digestive system cancers cause major public health problems and lead to high mortality rate worldwide \cite{song2019targeting}. Colorectal cancer and gastric cancer are the leading cause of digestive cancer mortality according to International
Agency for Research on Cancer and American Cancer Society \cite{bray2018global,siegel2019cancer}.

\textbf{{Motivation}.}
It is evident that the early stage diagnosis and treatment will significantly increase treatment success and thus reduce the mortality rate \cite{nazeri2018two}. Pathological checking is the golden standard for diagnosing these digestive system cancers. Generally, the pathological glass slides are made by the materials obtained in the operating room which are processed by formalin. To make the nuclei and cytoplasm visible, the slides are then dyed with hematoxylin and eosin (H \& E) \cite{guo2018breast}. During diagnosing phase, the specialists examine the glass slides under a microscope directly or check the generated digital pathology, such as the high-resolution whole slide image (WSI). The digital pathology based examination of WSI is becoming increasingly popular in recent years. Based on the observed features of the tissues, the pathological diagnostic results are then formed. 

However, pathological diagnosis is subjective with a high inter-rater variance \cite{gopinath2015development}. Besides, the experienced pathologists who are qualified for accurate diagnosis based on WSIs are scarce, and manual analysis of WSI is a time-consuming task for the pathologists due to the large size of WSI (e.g. 100000 $\times$ 100000) \cite{challenge-grand-2019}. Thus, an automatic and objective pathological WSI diagnostic model can be valuable to achieve early cancer detection and diagnosis.

\textbf{{Related Work.}}
A variety of approaches have been developed to conduct automatic diagnosis based on pathological WSIs \cite{huang2011time,mercan2016localization,roullier2010graph,cruz2014automatic}. Due to the large size of the WSI, the direct use of the entire image as the input of the machine learning algorithms is impossible because of the great memory usage requirement \cite{samsi2012efficient}. Related solutions include, downsampling and region of interest (RoI) detection \cite{huang2011time,mercan2016localization}, multi-resolution analyzing \cite{roullier2010graph}, and extracting image patches \cite{cruz2014automatic,korbar2017deep,sharma2017deep,mejbri2019deep}. 

To alleviate the computing complexity, Huang \textit{et al}. \cite{huang2011time} downsampled WSIs first and then detected the RoI at the low-resolution level. The authors in work \cite{mercan2016localization} proposed another diagnostically relevant RoIs location approach based on color and texture features. The produced probability maps can achieve 74\% overlap with the actual regions at which pathologists looked. Roullier \textit{et al}. proposed a highly efficient graph-based multi-resolution approach for mitosis extraction in breast cancer histological WSIs \cite{roullier2010graph}. They processed each resolution level with the focus of attention resulting from a coarser resolution level analysis, and the proposed segmentation was fully unsupervised by just considering domain-specific knowledge. 

The above downsampling and multi-resolution methods can alleviate the computing complexity through coarse WSI generation. However, this kind of method will introduce information loss due to the utilized coarse WSI, and thus ruin the diagnosis results. To solve this problem, many works perform patch splitting method to process WSI \cite{cruz2014automatic,tao2019highly,korbar2017deep,sharma2017deep}, of which the WSI is first divided into patches and then processed by the automatic algorithms one by one. Cruz-Roa \textit{et al}. \cite{cruz2014automatic} first cropped the WSIs into non-overlapping image patches of $100 \times100$ pixels via grid sampling. Then the author proposed to use a 3-layer CNN architecture to classify the extracted patches, and based on all the patch classification results to generate the final probability map for each WSI. To improve the analysis performance, the authors in work \cite{korbar2017deep} and work \cite{sharma2017deep} proposed to adopt larger patch sizes and use more complex CNN models to recognize each patch. However, it is inaccurate to just combine the patch-level classification results to analyze the whole WSI tissue. To further improve the result, semantic segmentation should be performed for each patch \cite{tao2019highly}.

\begin{figure}[h]
\centering
\includegraphics[width=3.49in]{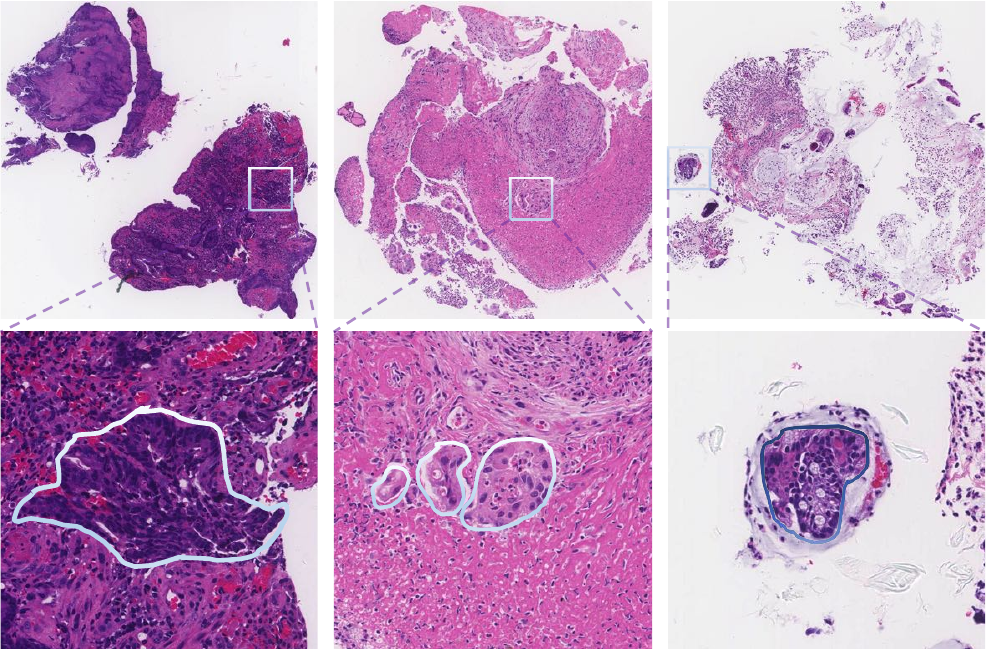}
\caption{Visualization of selected lesion patches from 3 different WSIs. These patches have high variations in appearance, such as lesion structure and stain style.}
\label{fig:Introduction}
\end{figure}

Traditional semantic segmentation methods \cite{ilea2011image}, \cite{preetha2012image} conduct image segmentation based on the hand-craft features. Although these methods can achieve satisfactory performance, the design of the hand-crafted features is based on complex domain knowledge and the ability of the segmentation model is insufficient. In the past several years, Fully Convolutional Networks (FCN) based method \cite{long2015fully} was proved can achieve decent semantic segmentation by modifying fully connected layers into convolution layers in CNN. The other improved state-of-the-art FCNs, such as U-Net \cite{ronneberger2015u,mejbri2019deep} and SkipDeconv-Net (SD-Net) \cite{roy2017error}, have shown great power in segmentation tasks. {According to the specific requirements of different segmentation tasks, the recent semantic segmentation works, such as \cite{fu2019stacked}, \cite{yin2019pm} and \cite{wang2019dual}, try to further improve segmentation performance using the stacked deconvolutional network (SDN), pyramid multi-label network (PM-Net), and dual encoding U-Net (DEU-Net), respectively.} However, automated segmentation of malignant lesion is very challenging due to high variations in appearance, especially when the patches are extracted from different WSIs scanned with different equipment or parameters, as shown in Fig. \ref{fig:Introduction}. Thus, the direct use of FCNs to conduct segmentation is insufficient.

Generally, the above appearance variations mean the dataset contains images with different data distributions. Most of the existing works address appearance variations by domain adaptation (DA) technique which treats the different data distributions as different domains \cite{pan2010survey}. {Many DA studies \cite{tsai2019domain,zou2019confidence} are performed on pixel-level road-scene semantic segmentation, such as the synthetic-to-real (GTA5 \cite{richter2016playing} to Cityscapes \cite{cordts2016cityscapes}).} In medical image processing, there are two kinds of DA solutions: pre-processing and domain-adversarial networks. The first kind of method, such as work \cite{Macenko2009A} which normalized the stain while retaining the structure, and work \cite{bentaieb2018adversarial} which proposed a discriminative image analysis model for stain standardization, just can alleviate the stain difference of the input image. To achieve more robust DA performance, many works propose to use domain-adversarial networks to impose constraints on the backbone network, and thus the backbone network can learn domain-invariant features \cite{lafarge2017domain,yang2018Generalizing,dou2018unsupervised,kamnitsas2017unsupervised}. Lafarge \textit{et al.} proposed a method based on domain-adversarial networks to remove the domain information from the model \cite{lafarge2017domain}. Yang \textit{et al.} proposed a novel online adversarial appearance conversion solution to explore a composite appearance and structure constraints \cite{yang2018Generalizing}. Dou \textit{et al.} proposed an unsupervised domain adaptation framework with a domain adaptation module (DAM) and a domain critic module (DCM) \cite{dou2018unsupervised}. However, the above methods only learn a single layer's domain-invariant feature, such as the last layer of the backbone network, and many feature maps of the network are ignored. The recent work \cite{kamnitsas2017unsupervised} concatenates the multi-layer cropped feature maps and passes them to a domain-adversarial discriminator. However, it is not conducive to the discriminator for classification due to the huge number of channels and the information loss introduced by feature cropping. {The recent researches conduct DA on different types of data, such as from the WSIs to microscopy images (MSIs) \cite{zhang2019whole}, or use pseudo-labeling for cross-modality microscopy image \cite{xing2019adversarial}. Most of these studies try to achieve unsupervised domain adaptation where the ground truth labels of the target domain are hard to obtain. In this paper, we focus on the supervised (the target domain labels are available) domain adaptation within the same data type (from WSIs to WSIs) but with different styles, such as the lesion structure and stain style.}


\textbf{Problems.} Two related problems are denoted as follows.

\textit{Problem1:} The existing directly applying patch-level segmentation to each WSI suffers the risk of false positive area detection. The WSIs which do not contain any malignant areas should be discarded before performing fine-grained segmentation. After WSI-level classification, key patches should be further selected from the malignant WSI with the similar reason: the patch without any malignant areas should not be processed by the segmentation model at all.
Besides, to train a patch-level classification model, a set of training patches with ground truth labels should be generated first. However, directly label each patch containing malignant tissue as positive sample is inaccurate because different patches contain different sizes of malignant areas.  

\textit{Problem2:} Given the selected key patches, how to design an appearance invariant image segmentation is still very challenging. The existing segmentation model, such as UNet, is lack of the ability of appearance invariant. The recent DA solutions are suffered from feature information loss due to cropping. 

The \textit{Digestive-System Pathological Detection and Segmentation Challenge 2019} (DigestPath2019), which is part of the \textit{MICCAI 2019 Grand Pathology Challenge}, set up a task for evaluating automatic algorithms on colonoscopy tissue screening from digestive system pathological images \cite{challenge-grand-2019}. This challenge provides a good platform to verify the above two issues. Besides, to conduct fair competition, the challenge requires each algorithm to execute on a single GPU and the average execution time on the test set 
can not exceed 120 seconds. This requires the designed scheme should take both accuracy and complexity into consideration.

\textbf{Approach and Contributions.}
In this paper, we proposed a multi-level colonoscopy malignant tissue detection incorporated with domain adaptive segmentation scheme. The multi-level architecture is adopted to realize the
malignant tissue detection in a coarse to fine manner, achieving lowering the risk of false positive detection while alleviating the high computing complexity at the same time. The domain adaptive segmentation is built to address the problem of appearance variations and thus boost the segmentation performance. We evaluated our method on the challenge dataset of the MICCAI 2019 challenge on lesion segmentation. Experimental results showed that our algorithm can achieve better result than other competitors. The main contributions of this paper are summarized as follows:

\begin{itemize}
\item We proposed a highly efficient multi-level malignant tissue detection architecture. In our architecture, the WSI-level classification is performed based on a patch-level classifier with a pre-prediction scheme. The WSIs without any malignant areas are dropped and thus the computing time is saved. For the selected positive WSIs, multiple patch-level models are trained with skillfully selected samples and then integrated together to choose the key patches. Besides, a malignant area ratio guided label smoothing scheme is applied to further increase the model accuracy.

\item We proposed an adversarial context-aware and appearance consistency (CAC-UNet) model to achieve robust appearance-invariant segmentation. Mirror designed discriminators are able to seamlessly fuse the whole feature maps of the generator without any information loss. The mask prior is further added to guide the accurate segmentation mask prediction through an extra mask-domain discriminator. Besides, several powerful strategies are integrated into the backbone of CAC-UNet to further improve the appearance-invariant ability.

\item The proposed scheme achieved the highest dice similarity coefficient (DSC) and area under the curve (AUC) score on the dataset of MICCAI 2019 challenge on colonoscopy tissue segmentation and classification task.

\end{itemize}

\textbf{Outline.} The paper is organized as follows. In Section \ref{sec:method}, we present the proposed multi-level lesion detection architecture and the domain adaptive segmentation model. Then, Section \ref{sec:experiment} reports the implementation, and analyses the experimental results. Finally, the discussion and conclusion of this paper are summarized in Section \ref{sec:discussion} and Section \ref{sec:conclusion}.

\section{Preliminaries}
\label{sec:Pre}
In this section, we clarify some related concepts and definitions, and introduce the basic knowledge about generative adversarial networks (GAN) that is important to the proposed method.

\textbf{Definitions and Concepts.} It is known that a WSI is very huge, and generally it is first cropped into patches for further processing. In this paper, the boldface uppercase letter $\mathbf{X}$ denotes a WSI, and boldface lowercase letter $\mathbf{x}$ denotes a patch. The boldface lowercase letter $\mathbf{y}$ and $\hat{\mathbf{y}}$ denote the predicted segmentation mask and ground truth mask for patch $\mathbf{x}$, respectively. Besides, the boldface uppercase letter $\mathbf{Y}$ and $\hat{\mathbf{Y}}$ denote a set of predicted and ground truth segmentation masks. The WSIs or patches that contain malignant area (with ground truth label 1) are denoted as positive WSIs or patches, otherwise they are denoted as negative samples. 

A domain refers to a dataset with a specific distribution, and different domains generally have images with different texture and appearances. According to work \cite{pan2010survey}, a domain $\mathcal{D}$ consists of two components: a feature space $\mathcal{X}$ and a marginal probability distribution $P(\mathbf{X})$. 


\textbf{GAN.} The framework of GAN is proposed for estimating generative models via an adversarial process \cite{goodfellow2014generative}. The target is to learn the generator's distribution $p_g$ over data $\textbf{x}$. To achieve this, in the GAN model a generator $G$ and a discriminator $D$ are defined. For a prior on input noise variables $p_z(z)$, the generator maps the variable $z$ to a generated data space $G(z)$. $D(\textbf{x})$ denotes that $\textbf{x}$ come from the real data rather than the generated one. The discriminator is trained targeting to tell apart real from fake input data and the generator is optimized to generate input data from the noise that “fools” the discriminator \cite{bentaieb2017adversarial}. Through the adversarial training mechanism, both the discriminator and the generator are then optimized. To summarize, generator $G$ and discriminator $D$ play the two-player minimax game with value function $V(G,D)$,

\begin{equation}
\min_{G}\max_{D}V(D,G)=\mathbb{E}_{\mathbf{x}\sim p_{\text{data}}(\mathbf{x})}\text{log}(D(\mathbf{x}))+\mathbb{E}_{z\sim p_{z}(z)}\text{log}(1-D(G({z})))
\label{equ:lde}
\end{equation}

\section{Proposed Method}
\label{sec:method}
In this section, we will first give the proposed multi-level detection architecture, and then introduce the WSI-level classification based on a patch-level model. After that, we will talk about the key patch selection and briefly explain the motivation of re-training the patch-level model in this stage. Based on the selected key patches, the domain adaptive segmentation will be discussed in detail to achieve the finest level detection.

\subsection{Multi-level detection architecture}
\begin{figure}[htb]
\centering
\includegraphics[width=3.48in]{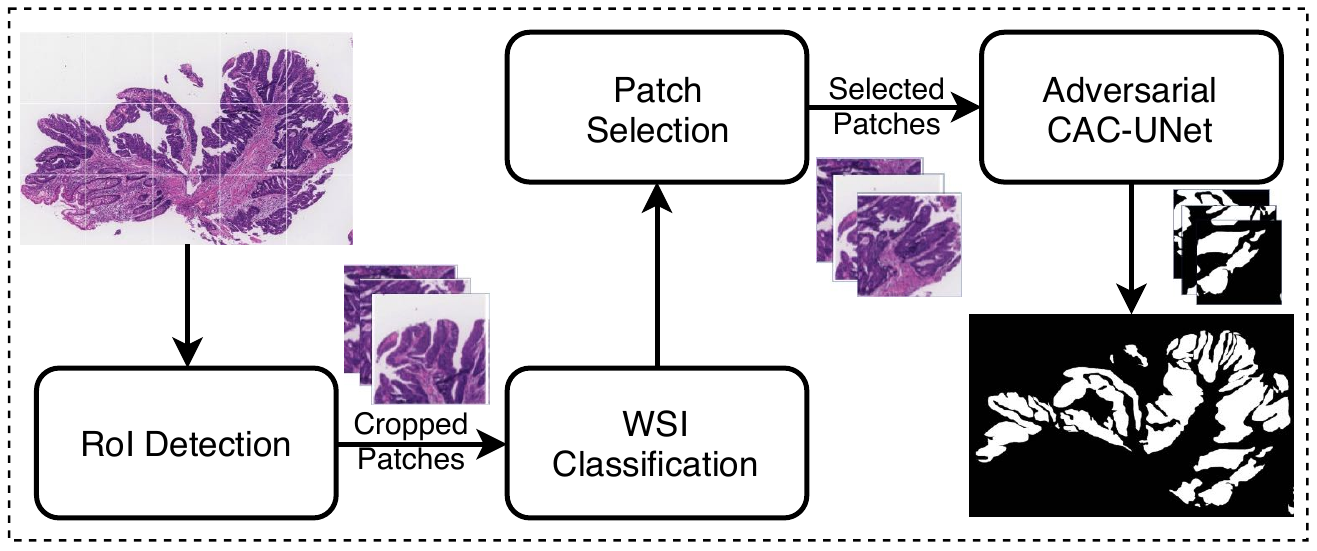}
\caption{An overview of the proposed WSI automatic multi-level detection architecture. A WSI is first processed by RoI detection module, and thus the background or the other unimportant areas are removed. For the detected RoI area, image cropping is performed and a series of patches are produced. The cropped patches are recognized and combined to conduct WSI-level classification. The selected positive WSI is then cropped and recognized again to choose the key patches. Finally, the selected positive patches are processed by adversarial CAC-UNet.}
\label{fig:arc}
\end{figure} 

The architecture of the proposed WSI automatic processing system is schematized in Fig. \ref{fig:arc}. The proposed multi-level architecture includes three main stages: \textbf{Stage-1} WSI-level classification, judging whether the input WSI is benign or malignant, and discarding the negative WSIs; \textbf{Stage-2} Key patch selection, finely classifying each patch in the malignant WSIs and choosing the positive ones as the key patches; \textbf{Stage-3} Segmenting the key patches and stitching them into a complete WSI mask. We adopt DenseNet \cite{huang2017densely} model for WSI-level classification and multi-model voting for patch-level classification. For the patch segmentation, we designed adversarial CAC-UNet to realize high accuracy segmentation.

\subsection{WSI classification}
\begin{figure}[htb]
\centering
\includegraphics[width=3.51in]{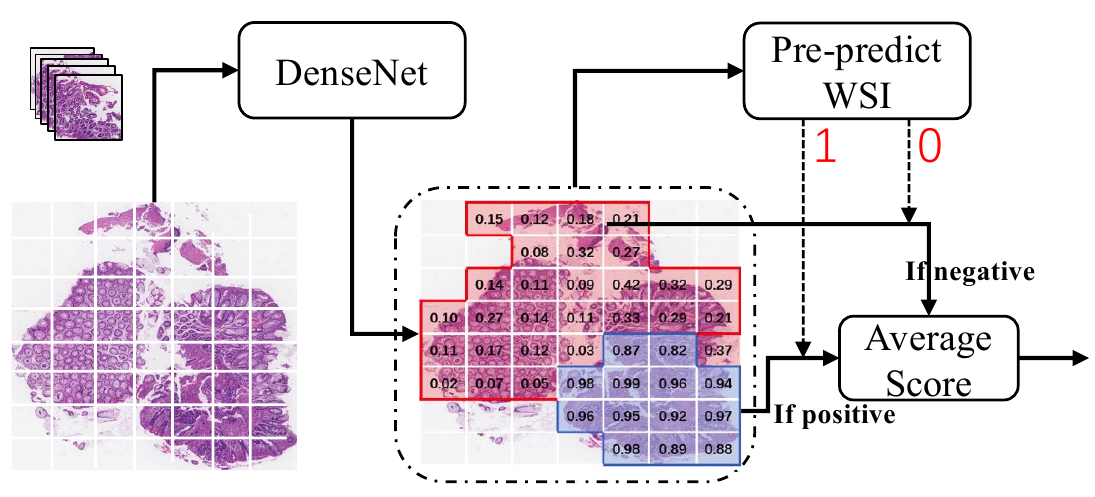}
\caption{WSI-level classification flow. The cropped patches are processed by DenseNet and the predicted probability map is generated. Guided by the WSI pre-prediction, the final WSI-classification score is produced by averaging selected patch-level results.}
\label{fig:WSI-classification}
\end{figure}

In this part, we propose to use a patch-level model to conduct WSI classification. We first classify all the patches cropped from the important areas of WSI, and get a set of classification results. If the patch classified as positive accounts for more than a certain percentage of all patches, we infer that the WSI is positive. We then use the average of all positive or negative patches' scores as the score for this WSI, as shown in Fig. \ref{fig:WSI-classification}.

\begin{figure}[htb]
\centering
\includegraphics[width=3.49in]{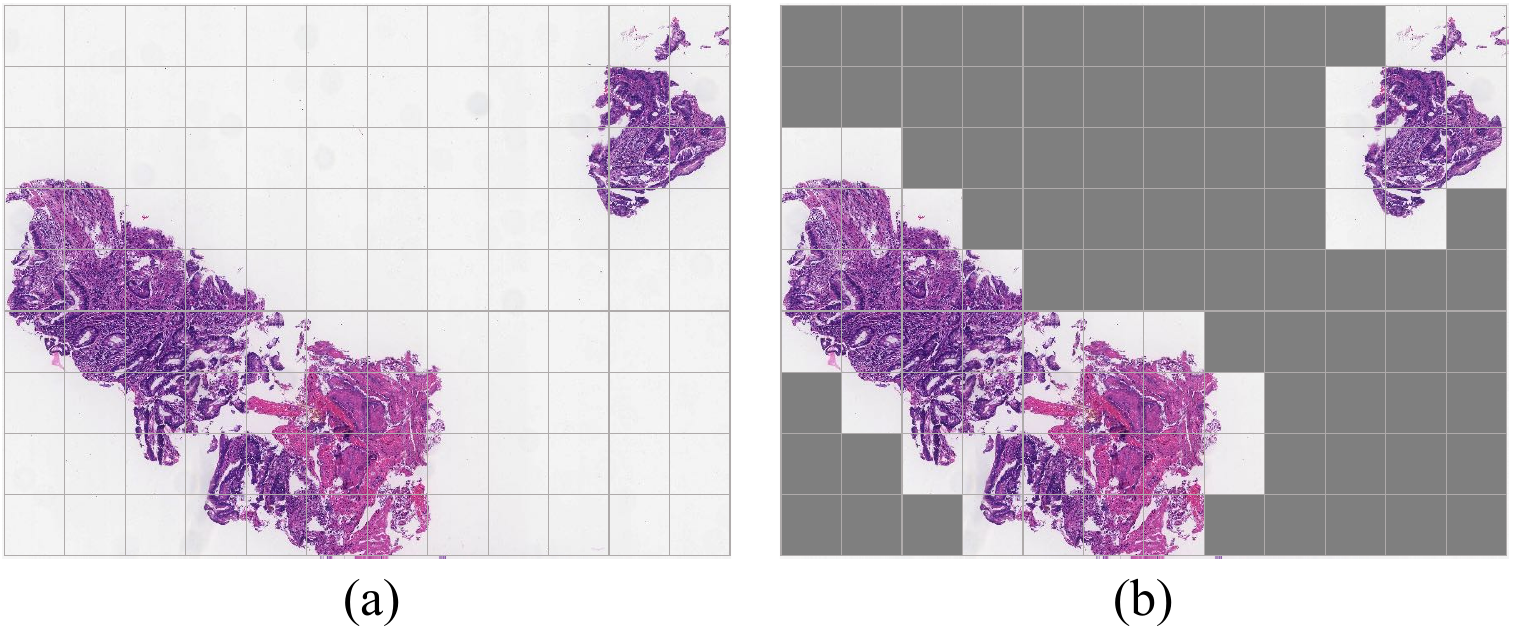}
\caption{Visualization of RoI for a selected WSI: (a) the original WSI, (b) the selected RoI areas.}
\label{fig:roi}
\end{figure}

Before performing WSI-level classification, we need to remove the irrelevant background areas. A simple patch based RoI detection is applied: if the standard deviation of RGB values is less than a pre-defined threshold $R$, the current patch will be discarded. After RoI detection, most part of the background area is removed, and a visualized RoI detection of a WSI is shown in Fig. \ref{fig:roi}.

Let $\mathbf{X}$ be a WSI after RoI detection, and $\mathbf{x}$ be a cropped patch from $\mathbf{X}$. We propose  $P(\mathbf{X})$ to perform WSI-level classification.

\begin{equation}
    P(\mathbf{X})=
    \begin{cases}
    Avg(\sum(P(\mathbf{x_i}))),\ \  \mathbf{x_i} \subset \mathbf{S_p}, & \text{if}\  \mathbf{X} \in {\hat{P}}; \\
    Avg(\sum(P(\mathbf{x_i}))),\ \  \mathbf{x_i} \subset \mathbf{S_n}, & \text{if}\  \mathbf{X} \in {\hat{N}}.
    \end{cases}
    \label{eq:WSI-prediction}
\end{equation}
where $\mathbf{S_p}$ and $\mathbf{S_n}$ are two patch-level sets, which include all the positive patches and negative patches of $\mathbf{X}$, respectively; $\hat{P}$ and $\hat{N}$ represent the pre-predicted positive and negative label, respectively; $Avg(\bullet)$ denotes the average function. We use DenseNet-161 with ImageNet pre-trained parameters \cite{huang2017densely} as the classifier to decide whether a patch $\mathbf{x}$ belongs to $\mathbf{S_p}$ or $\mathbf{S_n}$. Based on the classification results of all the patches, we then pre-predict whether $\mathbf{X} \in {\hat{P}}$ or $\mathbf{X} \in {\hat{N}}$. Then we can obtain the WSI classification score according to (\ref{eq:WSI-prediction}). Specifically, $\mathbf{x}$ is decided as belonging to $\mathbf{S_p}$, if $P(\mathbf{x}) \geq$ $\tau$ ($\tau$ is a threshold with constant value). Similarly, $\mathbf{x}$ belongs to $\mathbf{S_n}$, if $P(\mathbf{x}) <$ $\tau$. Then, we introduce (\ref{eq:pre-prediction}) to pre-predict the label of $\mathbf{X}$. Note that the predicted label by (\ref{eq:pre-prediction}) is a temporary intermediate result used to assist the generation of WSI classification score; the final WSI classification score is produced by (\ref{eq:WSI-prediction}).

\begin{equation}
    \begin{cases}
    \mathbf{X} \in {\hat{P}}, & \text{if}\  \frac{N_{\mathbf{S_p}}}{N_{\mathbf{S_p}} + N_{\mathbf{S_n}}} \geq T; \\
    \mathbf{X} \in {\hat{N}}, & \text{if}\  \frac{N_{\mathbf{S_p}}}{N_{\mathbf{S_p}} + N_{\mathbf{S_n}}} < T.
    \end{cases}
    \label{eq:pre-prediction}
\end{equation}
where $N_{\mathbf{S_p}}$ and $N_{\mathbf{S_n}}$ denote the patch number in set $\mathbf{S_p}$ and $\mathbf{S_n}$, respectively; $T$ is a threshold. In the following, we detail the training of our patch model adopted in this stage.

\begin{figure}[htb]
\centering
\includegraphics[width=1.31in]{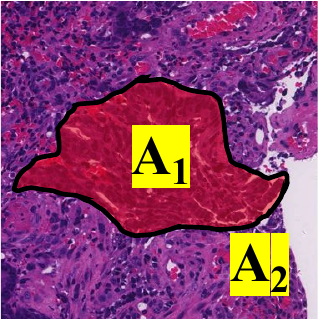}
\caption{Illustration of different areas of a patch: $A_1$ and $A_2$ represents the malignant and benign areas, respectively.}
\label{fig:area-vis}
\end{figure}

To train the patch-level model used in this stage, we first use a sliding window (stride=512, size=$1536 \times 1536$) to crop WSI images. We then perform online data augmentations to these cropped patches, which include random folds, random brightness contrast, and grid distortion. We sampled 50\% positive patches from positive WSIs and 50\% negative patches from negative WSIs as training data. Due to the cropping, the malignant area varies in different patches. Directly assign each patch containing malignant area with label 1 is unreasonable. To address this problem, we used label smoothing as introduced in work \cite{szegedy2016rethinking}. For the training example with ground-truth label y, the original label distribution $l_d$ is denoted as

\begin{equation}
l_d(k|\mathbf{x})= \delta_{k,y} 
\end{equation}
where $\mathbf{x}$ and $k$ are the training example and the corresponding label (malignant: $k = 1$, benign: $k = 0$); $\delta_{k,y}$ is Dirac delta, which equals 1 for k = y and 0 otherwise. Similarly, we replace the original label distribution as

\begin{equation}
\hat{l_d}(k|\mathbf{x})= (1-\epsilon)\delta_{k,y} + \epsilon a(k)
\end{equation}
where $\epsilon$ is a smoothing parameter and $a(k)$ is a distribution over labels. {Different with work \cite{szegedy2016rethinking} which selected uniform distribution for $a(k)$, in this paper we choose $a(k)$ as} 

\begin{equation}
a(k)= 
    \begin{cases}
    1-\frac{A_1}{A_1^{max}}, & \text{if}\  k = 0; \\
    \frac{A_1}{A_1^{max}}, & \text{if}\  k = 1.
    \end{cases}
\end{equation}
{where $A_1$ (as shown in Fig. \ref{fig:area-vis}) and $A_1^{max}$ are the malignant area of current patch and maximum malignant area of all the patches; $\frac{A_1}{A_1^{max}}$ represents the ratio of the malignant area in a patch to the maximum malignant area of all patches.}

Thus, in our work, the label distribution is written as

\begin{equation}
\hat{l_d}(k|\mathbf{x})= 
\begin{cases}
    (1-\epsilon)\delta_{k,y}+\epsilon(1-\frac{A_1}{A_1^{max}}), & \text{if}\  k = 0; \\
    (1-\epsilon)\delta_{k,y}+\epsilon\frac{A_1}{A_1^{max}}, & \text{if}\  k = 1.
    \end{cases}
    \label{eq:lsd}
\end{equation}

{Based on (\ref{eq:lsd}), we will change the ground-truth label distribution according to the malignant ratio $\frac{A_1}{A_1^{max}}$. Take an malignant patch (y = 1) for example, if this patch has big $\frac{A_1}{A_1^{max}}$, we will encourage it to be confident with the ground truth label y; if the patch has small $\frac{A_1}{A_1^{max}}$, which means the malignant patch having many benign areas, thus we should encourage it to be less confident with the ground truth label y.}


\subsection{Key Patch Selection}
In this stage (Stage-2), the target is to find all the positive patches in the WSI if it is judged as positive. Note that we do not directly use the patch model trained in Stage-1. 

\begin{figure}[htb]
\centering
\includegraphics[width=3.51in]{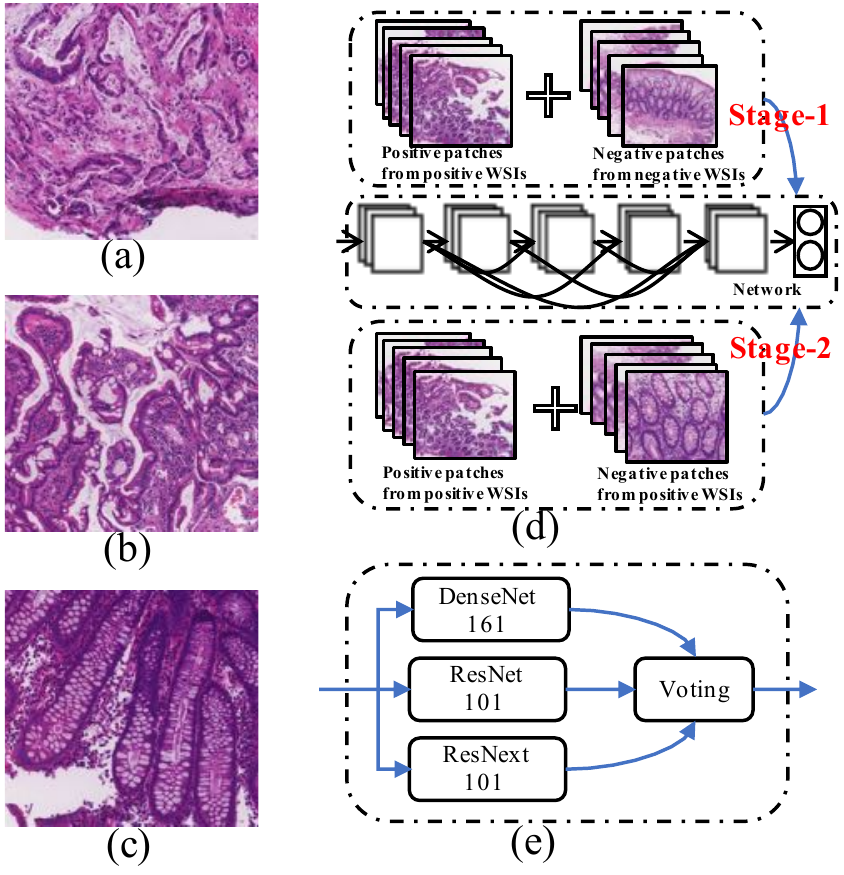}
\caption{(a) a malignant patch from positive WSI, (b) a benign patch from positive WSI, (c) a benign patch from negative WSI, (d) comparison of different training strategies for Stage-1 and Stage-2, (e) multi-model voting scheme.}
\label{fig:ps}
\end{figure}

In Fig. \ref{fig:ps} (a) to Fig. \ref{fig:ps} (c), three patches are visualized: a positive patch from the positive WSI, a negative patch from the positive WSI, and a patch from the negative WSI. As denoted by the figure, the glandular structure of the malignancy patch, Fig. \ref{fig:ps} (a), appears in heterogeneous shapes, but the benign patch, Fig. \ref{fig:ps} (c), have a typical and uniform glandular arrangement. However, we should note that the glandular structure of the benign patch from the positive WSI, Fig.  \ref{fig:ps} (b), seems very different from the benign patch from the negative WSI, Fig. \ref{fig:ps} (c). In fact, the benign patches from the negative WSI are more like normal lesion. In Stage-2, our target is to extract all the malignancy patches like Fig. \ref{fig:ps} (a) from the positive WSIs, and the patches like Fig. \ref{fig:ps} (b) can not provide useful information to assist the classification. Thus, the patches from negative WSIs are not used for training in this stage. We remind that in Stage-1, our aim is to distinguish the positive WSIs from the negative WSIs based on a patch-level model. In order to achieve more effective classification, the positive patches from the positive WSIs and patches from the negative WSIs are selected as the important information for positive WSIs and negative WSIs, respectively. The comparison of training strategies for Stage-1 and Stage-2 are summarized as Fig. \ref{fig:ps} (d).

In work \cite{zhu2019breast}, the authors proposed to assemble multiple hybrid models with the same architecture to reduce generalization error and improve performance. Different from work \cite{zhu2019breast}, we apply different state-of-the-art CNN models predicting together. DenseNet connects each layer to every other layer in a feed-forward fashion, and thus it can be deeper and more accurate \cite{huang2017densely}. ResNext \cite{xie2017aggregated} is able to improve the classification accuracy by increasing the number of repeated basic building block that aggregates a set of transformations with the same topology. ResNet adopts a residual learning framework to ease the training of networks, which can make the networks substantially deeper than the previous models \cite{he2016deep}.
These three models are designed based on different ideas and they have some degree of complementary features. Based on these three models, we perform multi-model voting scheme to conduct the final classification by averaging the predicted scores of different models, as shown in Fig. \ref{fig:ps} (e). DenseNet161, ResNet101, and ResNext101 are chosen in this work.

\subsection{Segmentation}
Our goal is to extract the pixel-level segmentation mask for the selected key patches and thus build the whole lesion segmenting result for the WSI. In this part, we first present related definitions and then give our proposed adversarial CAC-UNet architecture. After that, the backbone of CAC-UNet and the adversarial learning are detailed.

\subsubsection{Definitions in Segmentation}
To learn domain-invariant features and increase the generalization ability of our model, we split the training set $\mathbf{X}$ into $\mathbf{X}_A$ and $\mathbf{X}_B$ subsets through clustering according to different texture and appearances of the WSIs, such as gland structure, stain style, and lesion distribution. For convenience, we use $\mathcal{D}_A$ and $\mathcal{D}_B$ representing two domains corresponding to subset $\mathbf{X}_A$ and $\mathbf{X}_B$. We further define another two domains $\mathcal{D}_{Pmask}$ and $\mathcal{D}_{Gmask}$ to denote the model predicted segmentation masks and the expert labeled ground truth masks. We make a hypothesis that the expert labeled ground truth is often with a smooth and continuous boundary. If we put this prior constraint into the model training, the accuracy of the predicted masks will thus be improved.  

To realize the domain-invariant feature learning, our model is built on the foundations of GAN. 
In our work, the discriminators refer to a series of classifiers based on CNNs, and the generator means the entire or part of the segmentation model, which generates some feature maps or segmenting masks as the input to different discriminators.

\subsubsection{Architecture of the Proposed Adversarial CAC-UNet}

\begin{figure}[htb]
\centering
\includegraphics[width=3.49in]{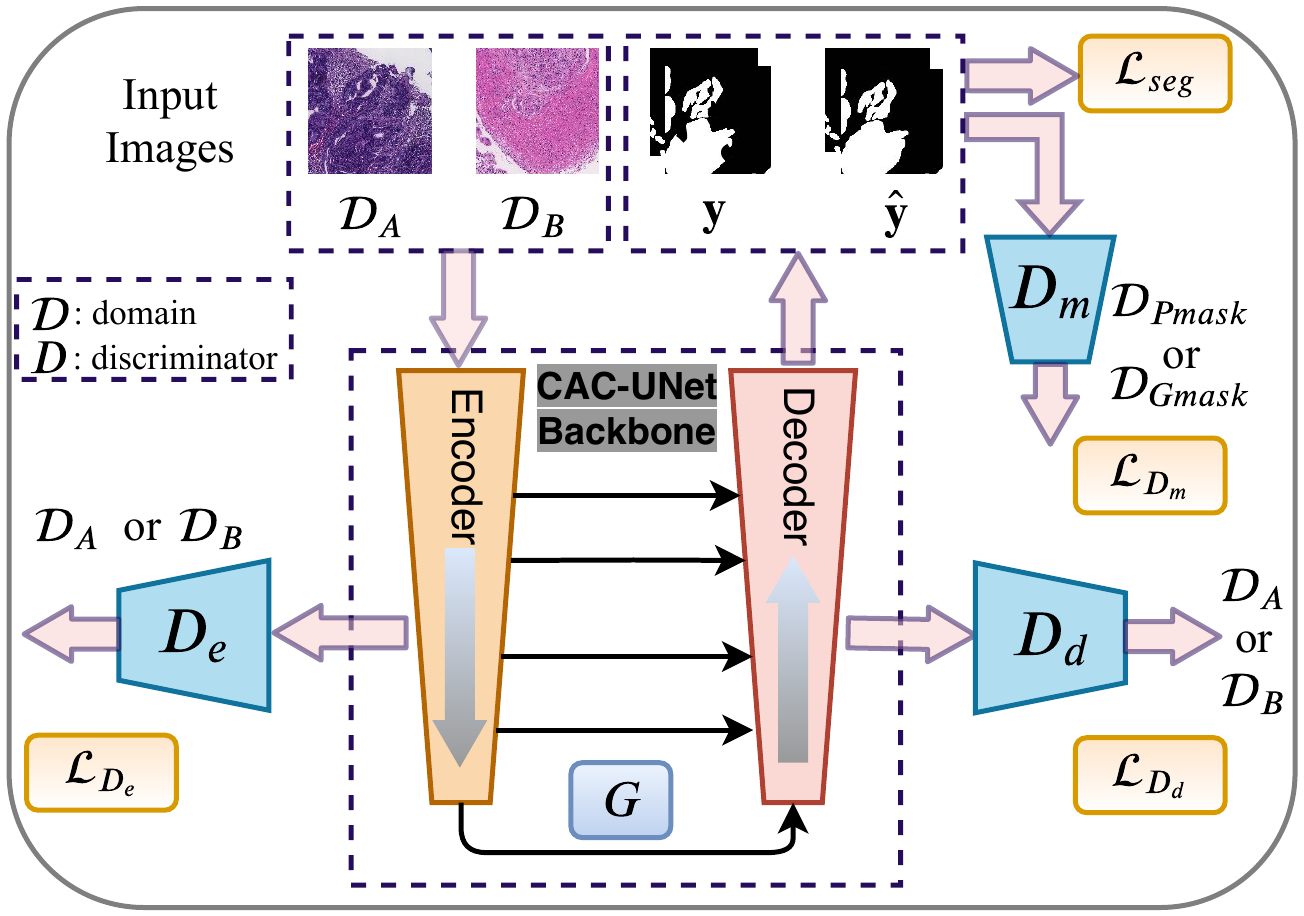}
\caption{Overview of our proposed adversarial CAC-UNet framework. The backbone of CAC-UNet is the generator and serves the image segmentation. The discriminators {$De$,$Dd$,$Dm$} differentiate their inputs accordingly and thus generate adversarial losses.}
\label{fig:top-seg}
\end{figure}

In Fig. \ref{fig:top-seg}, the architecture of the proposed adversarial CAC-UNet is presented. Our architecture is composed of the main segmentation network, backbone of CAC-UNet, and three discriminators, $D_e$, $D_d$ and $D_m$. Note that the backbone of segmentation network CAC-UNet plays the role of generator $G$.

The backbone of CAC-UNet is constructed based on the basic UNet \cite{ronneberger2015u}, and consists of an encoder and a decoder part. The encoder takes the image as the input and maps it into feature maps. The decoder takes these feature maps as the input and transforms them to segmentation mask, which will be compared with the ground truth mask. The discriminator $D_e$ takes the encoder feature maps as the input and then combines them with the feature maps generated by $D_e$ itself to decide whether the input image belongs to domain $\mathcal{D}_A$ or $\mathcal{D}_B$. Similarly, the discriminator $D_d$ combines the decoder feature maps and the feature maps generated by $D_d$ to conduct the same decision process. The discriminator $D_m$ takes both the predicted segmentation mask and the ground truth mask as the input to recognize whether the mask is generated by the model ($\mathcal{D}_{Pmask}$ domain) or the expert ($\mathcal{D}_{Gmask}$ domain).  

In summarize, the optimization targets of our model are to: (1) Minimize the differences between the predicted segmentation masks and the ground truth segmentation masks; (2) discriminate images from domain $\mathcal{D}_A$ from domain $\mathcal{D}_B$ based on the encoder feature maps; (3) discriminate images from domain $\mathcal{D}_A$ from domain $\mathcal{D}_B$ based on the decoder feature maps; (4) discriminate the predicted masks from ground truth masks. To achieve these four targets, the proposed final adversarial training optimization loss function is denoted as 

\begin{equation}
    \mathcal{L}_{full}=\mathcal{L}_{seg}+\alpha_{e} \mathcal{L}_{D_{e}^{adv}}+\alpha_{d}\mathcal{L}_{D_{d}^{adv}}+\alpha_{m}\mathcal{L}_{D_{m}^{adv}}\label{equ:4}
\end{equation}
where $\alpha_{e}$, $\alpha_{d}$, $\alpha_{m}$ are trade-off parameters adjusting the importance of each term. 

\subsubsection{Backbone of CAC-UNet} 
In the basic UNet \cite{ronneberger2015u}, the skipped encoder feature maps and the up-sampled decoder feature maps are concatenated to perform segmentation by recovering the full spatial resolution at the model output. However, the disadvantages are obvious when processing histopathology images: the lack of the context aware ability and suffering context information loss, inability in handling the appearance inconsistency. We design the backbone of our CAC-UNet architecture targets on more powerful context aware and appearance invariant ability. The key details are highlighted as follows. 

\begin{figure}[h!]
\centering
\includegraphics[width=3.50in]{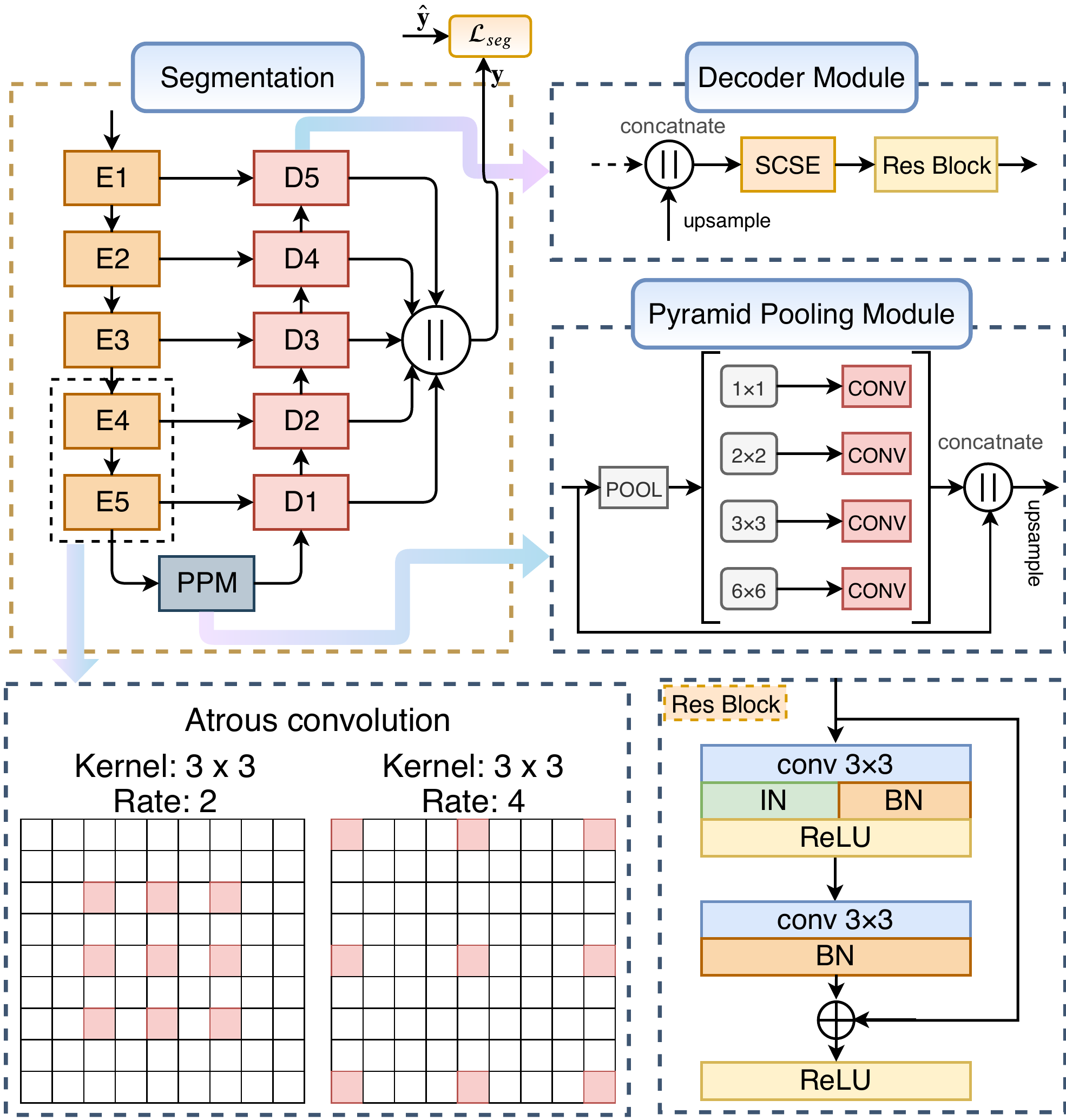}
\caption{The backbone structure of our CAC-UNet. Top left: backbone segmentation structure; top right: decoder module; middle: PPM; bottom left: atrous convolution; bottom right: Res Block with IN in the decoder module.}
\label{fig:structure}
\end{figure} 

\textbf{Encoder Selection.} 
We have tried different more powerful encoders. We chose the widely used ResNet and its improved ResNext as the encoder, and we tried models with different depths, such as 34, 50, 101. Through experiments, it is found that ResNet50 can achieve better performance on the validation set. While larger networks such as ResNet101 and ResNext101 have no obvious advantages, but there is a risk of over-fitting. Therefore, we choose ResNet50 as the encoder of the segmentation network. 

\textbf{Context-aware Design.} To enhance the context-aware ability, we implemented two techniques proposed by work \cite{roy2018concurrent} and \cite{zhao2017pyramid} into the UNet: the Spatial-Channel Sequeeze \& Excitation (SCSE) block and the Pyramid pooling module (PPM).

SCSE block is able to adjust the weighting of different network feature maps according to their importance: attach a higher weight to important feature maps or feature channels and attach small weight to reduces the influence of unimportant features. In our design, we integrate the SCSE attention block to the decoder part for realizing the context aware. 

In UNet, the input image is encoded as a multi-channel (1024) feature maps through the processing of the encoder, and then the decoder transforms the feature maps to the final segmenting mask by a series of up-sampling and skip connected feature concatenation. The encoded multi-channel feature map contains most part of the information used for segmentation. However, the single scale will inevitably introduce context information loss. Motivated by work \cite{zhao2017pyramid}, we put hierarchical PPM in the center of the network to aggregate more global information, which embeds information with different scales and varying among different sub-regions. Following the same structure with the PPM in \cite{zhao2017pyramid}, we also adapt it to fuse four different pyramid scale feature maps, as shown in the middle of Fig. \ref{fig:structure}. Note that the atrous convolution is used (kernel: $3 \times 3$, rate = 2, 4) in the encoder unit E4 and E5 when enabling PPM scheme, as shown in lower left part of the Fig. \ref{fig:structure}. 

\textbf{Appearance Consistency Design.}
To force our CAC-UNet to learn features that are invariant to appearance changes, such as stain colors, lesion structure styles, we add the instance normalization (IN) \cite{ulyanov2017improved} function to our network like the proposed IBN block by work \cite{pan2018two}.

As denoted by work \cite{pan2018two}, IN is stronger for learning appearance invariant features and batch normalization (BN) is essential for preserving content related information. Thus we also apply IN and BN at the same time. In the encoder part, we only integrate IBN in E2 to E4, in order to enhance the domain adaptability of the model. In the decoder part, we embed IN into all the residual blocks of the decoding units, D1 to D5. The detailed structure is depicted as the lower right part of the Fig. \ref{fig:structure}. 

\textbf{Feature Fusion.}
To achieve higher segmentation accuracy, we conduct pixel-level mask prediction based on hypercolumn \cite{hariharan2015hypercolumns}. The hypercolumn at a pixel is defined as the vector of activations of all feature map units at the same pixel-level location. Thus, the hypercolumn can help address the problem that: it is too coarse just considering the information of the decoder output. We upsampled the features of last layer in the decoding units and concatenated them to obtain a hypercolumn, which is used to predict the final segmentation mask. Through this scheme, the multi-scale features including both the global semantic information and the precise localization information are fused.

\textbf{Segmentation Loss Function.} To the segmentation output of CAC-UNet backbone, we apply the segmentation loss $\mathcal{L}_{seg}$ as

\begin{equation}
    \mathcal{L}_{seg}=Dice(\mathbf{x},\mathbf{y})
    \label{equ:loss-seg}
\end{equation}
where $Dice(\bullet)$ represents the Dice loss; $\mathbf{x}$ and $\mathbf{y}$ denotes an image patch and the corresponding segmentation mask, respectively.

\subsubsection{Domain-adversarial Learning}
\begin{figure}[htb]
\centering
\includegraphics[width=3.51in]{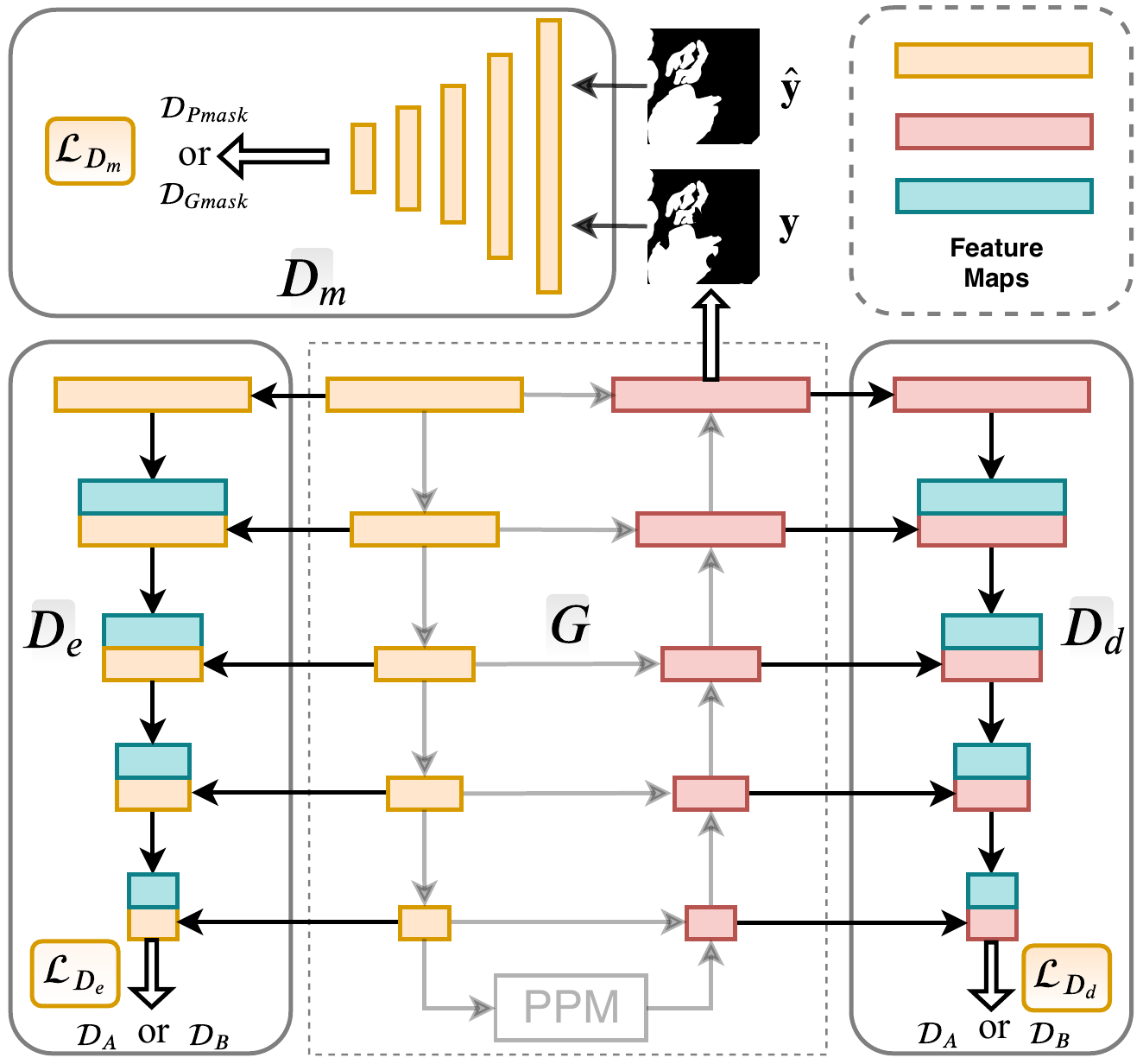}
\caption{The structures of three discriminator: $D_e$, $D_d$ and $D_m$. $D_e$ and $D_d$ are mirror models of the encoder and decoder, respectively. $D_m$ adopts the same structure with the encoder.}
\label{fig:discriminator}
\end{figure}

\textbf{Targets.} Target 1: learn feature maps that are invariant to different domains ($\mathcal{D}_A$ and $\mathcal{D}_B$), thus the segmentation network can robustly segment images from different domains. For this target, the encoder and decoder of the segmentation model are served as the generators ($G_e$ and $G_d$). Target2: Put mask prior to the model, and make the generated masks more like the ground truth. For this target, the whole segmentation model is served as the generator ($G_m$). To perform adversarial learning and realize the optimization of these generators, we need to design discriminators ($D_e$, $D_d$ and $D_m$) correspondingly, which will be detailed in the following. Although these discriminators will not be used in the inference stage, they are vital for adversarial learning and the proper discriminator can help the model achieve the above two targets efficiently. All the discriminators are presented in Fig. \ref{fig:discriminator}.

\textbf{Discriminator $D_e$ and $D_d$.} 
For discriminator $D_e$ and $D_d$, the first primary issue is to choose which layer’s feature maps in the segmentation network as the input of the discriminator. It is intuitive to select the feature maps of the last layer (such as the last layer of the encoder or decoder) because these feature maps contain more discriminative high-level semantic information. However, in \cite{kamnitsas2017unsupervised}, they found that it is not ideal to only select feature maps of the last layer to adapt because the early layers are more susceptible to appearance variations between domains. In order to ensure that all feature maps to be concatenated and adapted, the authors in \cite{kamnitsas2017unsupervised} crop large size feature maps to match the size of the last layer and then concatenate them. However, due to the cropping, a lot of information is lost. Moreover, the number of the directly concatenated feature maps is too huge, which is difficult for the discriminator training.

To avoid feature map information loss and decently adjust the weights of these features at the same time, we designed two mirror networks of the encoder and decoder as discriminator $D_e$ and $D_d$, as shown in Fig. \ref{fig:discriminator}: the left and the right part. Thus, our discriminator ($D_{e}$) uses a similar network structure with the encoder. The key details of discriminator ($D_{e}$) are denoted as follows: (1) The first layer of discriminator $D_{e}$ takes the first layer's feature maps of the encoder as the input directly; (2) The other layers of $D_{e}$ will generate same size feature maps as the the corresponding encoder layers, and then are sequentially concatenated to the feature maps from the corresponding encoder layer. The decoder discriminator ($D_{d}$) is constructed in the same manner. Through the proposed mirrored discriminators ($D_{e}$, $D_{d}$), we can ingeniously solve the problem of the inconsistent size of different layers' feature maps instead of cropping the feature maps roughly so that the discriminator is able to use different layers' feature maps completely without any loss.

We adopt binary cross entropy as loss to update parameters of $D_{e}$ or $D_{d}$. The losses of $D_{e}$ and $D_{d}$ are shown as Eq. (\ref{equ:lde}) and Eq. (\ref{equ:ldd}), 

\begin{equation}
\mathcal{L}_{D_{e}}=-\mathbb{E}_{{\mathbf{x}}\sim p_{B}({\mathbf{x}})}\text{log}(D_{e}(G_{e}({\mathbf{x}})))-\mathbb{E}_{{\mathbf{x}}\sim p_{A}({\mathbf{x}})}\text{log}(1-D_{e}(G_{e}({\mathbf{x}})))
\label{equ:lde}
\end{equation}

\begin{equation}
\mathcal{L}_{D_{d}}=-\mathbb{E}_{{\mathbf{x}}\sim p_{B}({\mathbf{x}})}\text{log}(D_{d}(G_{d}({\mathbf{x}})))-\mathbb{E}_{{\mathbf{x}}\sim p_{A}({\mathbf{x}})}\text{log}(1-D_{d}(G_{d}({\mathbf{x}})))
\label{equ:ldd}
\end{equation}
where $p_{B}(\mathbf{x})$ is the distribution of data $\mathcal{D}_B$, $p_{A}(\mathbf{x})$ is the data distribution of $\mathcal{D}_A$, $G_{e}(\mathbf{x})$ is the feature maps of encoder, and $G_{d}(\mathbf{x})$ is the feature maps of decoder.

During adversarial training, the losses of $D_{e}$ and $D_{d}$ are shown as Eq. (\ref{equ:lde_adv}) and Eq. (\ref{equ:ldd_adv}).
\begin{equation}
\mathcal{L}_{D_{e}^{adv}}=-\mathbb{E}_{{\mathbf{x}}\sim p_{A}({\mathbf{x}})}\text{log}(D_{e}(G_{e}({\mathbf{x}})))-\mathbb{E}_{{\mathbf{x}}\sim p_{B}({\mathbf{x}})}\text{log}(1-D_{e}(G_{e}({\mathbf{x}})))
\label{equ:lde_adv}
\end{equation}

\begin{equation}
\mathcal{L}_{D_{d}^{adv}}=-\mathbb{E}_{{\mathbf{x}}\sim p_{A}({\mathbf{x}})}\text{log}(D_{d}(G_{d}({\mathbf{x}})))-\mathbb{E}_{{\mathbf{x}}\sim p_{B}({\mathbf{x}})}\text{log}(1-D_{d}(G_{d}({\mathbf{x}})))
\label{equ:ldd_adv}
\end{equation}

\textbf{Discriminator $D_m$.} As we presented before, the expert labeled ground truth masks ($\hat{\mathbf{y}}$) are often with smooth and continuous boundaries. We use this prior to guide the prediction of the segmentation network, by introducing an additional mask loss. We adopt mask discriminator $D_m$ to achieve this. The structure of $D_m$ is depicted as the top part of Fig. \ref{fig:discriminator}.

It distinguishes the mask between the domain $\mathcal{D}_{Pmask}$ and the domain $\mathcal{D}_{Gmask}$. By adversarial learning, it can make the output of the segmentation network as close as possible to the ground truth, thus making their boundaries similar. We also adopt binary cross entropy as the loss to update the parameters of $D_{m}$, which is shown as Eq. (\ref{equ:ldm}).

\begin{equation}
    \mathcal{L}_{D_{m}}=-\mathbb{E}_{\hat{\mathbf{y}}\sim p_{Gmask}(\hat{\mathbf{y}})}\text{log}(D_{m}(\hat{\mathbf{y}}))-\mathbb{E}_{\mathbf{x}\sim p_{Gmask}(\mathbf{x})}\text{log}(1-D_{m}(G(\mathbf{x})))
    \label{equ:ldm}
\end{equation}

During adversarial training, we adopt Eq. (\ref{equ:ldm_adv}) as the loss of $D_{m}$.
\begin{equation}
    \mathcal{L}_{D_{m}^{adv}}=-\mathbb{E}_{\mathbf{x}\sim p_{Gmask}(\mathbf{x})}\text{log}(1-D_{m}(G(\mathbf{x})))
    \label{equ:ldm_adv}
\end{equation}

\textbf{Training.} With the images and labels in both $\mathcal{D}_{A}$ and $\mathcal{D}_{B}$, we can train the segmentation network ($G$) and the discriminators ($D_{e}$, $D_{d}$, $D_{m}$) in a supervised way. In the training phase, we try to make $G$ segment more accurately by adopting feature maps invariant to variations between $\mathcal{D}_{A}$ and $\mathcal{D}_{B}$. In the initial stage, we train $G$ with $\left ( \hat{\mathbf{X}}, \hat{\mathbf{Y}} \right )$ by minimizing $\mathcal{L}_{seg}$, where $\hat{\mathbf{X}}$ is the collection of image patches randomly sampled from $\mathcal{D}_{A}$ or $\mathcal{D}_{B}$, and $\hat{\mathbf{Y}}$ is the collection of their label masks. After training $G$ for $s_{0}$ epochs, we start to train $D_{e}$, $D_{d}$, $D_{m}$ independently for $d_{0}$ epochs with the trained $G$ by minimizing $\mathcal{L}_{D_{e}}$, $\mathcal{L}_{D_{d}}$ and $\mathcal{L}_{D_{m}}$. Then, we obtain a initial $G$ and initial $D_{e}$, $D_{d}$, $D_{m}$ which can initially classify, and we start adversarial training them alternately until convergence \cite{mei2020cross}. In particular, we use $\mathcal{L}_{full}$ as the loss of segmentation network instead of $\mathcal{L}_{seg}$ when training alternately.

\section{Experiment}
\label{sec:experiment}
\subsection{Dataset}
\label{sec:data}
We evaluate our method on colonoscopy tissue segment dataset of \textit{MICCAI 2019 Challenge} DigestPath2019 \cite{challenge-grand-2019}.
The training set contains a total of 450 patients' 750 tissue slices of an average size of $3000 \times 3000$. The fine pixel-level annotations of lesion and the diagnosis of the tissues are labeled by experienced pathologists. The testing set contains another 150 patients' 250 tissues. All WSIs were stained by hematoxylin and eosin and scanned at X20. Note that the testing set is not released to the public to guarantee that the test data cannot be included in the training procedure. To train and verify our model, we split the 750 WSIs into two parts: 682 WSIs for training and 68 WSIs for validation.

{Except for the DigestPath2019 dataset, we also employ another two pathology image datasets to help validate our technologies, such as the label smoothing and domain adaptation. The first dataset is built based on Camlyon16 dataset \cite{Camelyon2016}, including 50,000 training patches, 50,000 validation patches with size $1536 \times 1536$ cropped from the training set of Camlyon16, and 60 testing WSIs (37 normal and 23 tumor). The second dataset contains renal biopsy pathology images with size $1024 \times 1024$ from clinical routines of one top-tier hospital in Beijing, which are stained with Periodic Schiff-Methenamine Silver (PASM) or Periodic acid-Schiff (PAS) method. The training set consists of 6708 patches (4324 PASM stained patches and 2384 PAS stained patches), and the testing set includes 1524 patches (912 PASM stained patches and 612 PAS stained patches). The aim of the second dataset is to build model segmenting the glomeruli from the renal biopsy pathology images.}

\subsection{Implementation}
\label{sec:impl}
The proposed method for WSI classification is implemented with Python3.6 and Pytorch0.4.1 using an NVIDIA
GeForce GTX 1080 Ti GPU. All the training patches are cropped from 682 WSIs, using a patch size of $1536 \times 1536$ and a stride of 512 pixels. We trained networks with standard back propagation, which is performed by stochastic gradient descent method (momentum = 0.9 with weight decay 0.0001, batch size = 64, constant learning rate = 0.001), and the models converge to its optimal accuracy within 5 epochs. The proposed segmentation network is implemented with the Pytorch 1.0 framework with a Tesla V100. We used the RAdam and Lookahead optimizer (initial learning rate is 0.001, momentum parameters $\beta_1$ = 0.95, $\beta_2$ = 0.999, weight decay = 0.0005, batch size = 8) to update the parameters of the networks. All the training patches are cropped from 223 WSIs, using a patch size of $1536 \times 1536$ and a stride of 512 pixels. During training, these patches are resized to $512 \times 512$. We first trained the segmentation network for 20 epochs and fine-tune it with the domain-adversarial learning for 5 epochs.

The other related parameters of this work are summarized in Table \ref{tab:parameters}.
\begin{table}[]
	\centering
\begin{tabular}{@{}lllllll@{}}
\toprule
\textit{$R$} & $\tau$ & $T$ & $\epsilon$ & $\alpha_{e}$ & $\alpha_{d}$ & $\alpha_{m}$ \\ \midrule
         30    &     0.1   &   0.1  &    0.1   &    0.01  &      0.001      &    0.001          \\ \bottomrule
\end{tabular}
\caption{Other related parameters.}
    \label{tab:parameters}
\end{table}

\subsection{Evaluation Criteria}
The evaluation of the WSI classification and lesion segmentation follows the challenge rule\footnote{\url{https://digestpath2019.grand-challenge.org/Evaluation/}}. Classification accuracy, recall and precision are also involved in the evaluation of Stage-2 patch models. Besides, we will briefly analyze the computing complexity for our scheme. 

\textbf{WSI classification:} The WSI classification is evaluated by classification area under the curve (AUC). AUC is equal to the probability that a classifier will rank a randomly chosen positive instance higher than a randomly chosen negative one. AUC is denoted as

\begin{equation}
A=\int_{x=0}^{1} TPR\left(FPR^{-1}\left(x\right)\right)\,dx=P\left(X_{1}>X_{0}\right)
\label{equ:AUC}
\end{equation}
where $X_1$ and $X_0$ are the scores for a positive and a negative instance, respectively; $TPR$ represents true positive rate, and $FPR$ represents false positive rate.

\textbf{Accuracy, Recall and Precision.} Accuracy is the ratio of the corrected predicted images to the whole pool of validation samples. Recall is the proportion of real positives cases that are correctly predicted positive. Conversely, precision indicates the proportion of predicted positive cases that are correctly real positives. The three evaluation metrics above are depicted as follows:
\begin{equation}
Accuracy = \frac{N_{tp}+N_{tn}}{N_{tp}+N_{fp}+N_{tn}+N_{fn}}
\label{equ:acc}
\end{equation}

\begin{equation}
Recall = \frac{N_{tp}}{N_{tp}+N_{fn}}
\label{equ:recall}
\end{equation}

\begin{equation}
Precision = \frac{N_{tp}}{N_{tp}+N_{fp}}
\label{equ:precision}
\end{equation}
where $N_{tp}$,$N_{fp}$,$N_{tn}$ and $N_{fn}$ denote the number of true positives(TP), false positives(FP), true negatives(TN) and false negatives(FN) respectively.

\textbf{Lesion segmentation:} The lesion segmentation is evaluated by Dice Similarity Coefficient (DSC). The Dice metric measures area overlap between segmentation results and ground truth annotations. DSC can be written as 
\begin{equation}
Dice=\frac{2 |A\cap B|}{\left|A\right|+\left|B\right|}\times 100\%
\label{equ:DSC}
\end{equation}
where $A$ and $B$ denote the sets of foreground pixels in the annotation and the corresponding sets of foreground pixels in the predicted segmentation result, respectively.

\begin{table*}[!htbp]
\centering
\begin{tabular}{@{}cccccc@{}}
\toprule
Team            & DSC    & DSC Rank & AUC    & AUC Rank & Final Rank \\ \midrule
\textbf{kuanguang} & \textbf{0.8075} & \textbf{1} & \textbf{1.0000} & \textbf{1} & \textbf{1}         \\
zju\_realdoctor & 0.7789 & 5        & 1.0000 & 1        & 2          \\
TTA\_Lab        & 0.7878 & 3        & 0.9948 & 4        & 3          \\
SJTU\_MedicalCV & 0.7928 & 2        & 0.9773 & 6        & 4          \\
ustc\_czw       & 0.7862 & 4        & 0.9784 & 5        & 5          \\
chenpingjun     & 0.7197 & 8        & 0.9974 & 3        & 6          \\
MCPRL\_218      & 0.7397 & 7        & 0.9745 & 8        & 7          \\
path\_fitting   & 0.6794 & 10       & 0.9754 & 7        & 8          \\
mirl\_task2     & 0.7590 & 6        & 0.5164 & 13       & 9          \\
Roselia         & 0.6920 & 9        & 0.8886 & 11       & 10         \\ \bottomrule
\end{tabular}
    \caption{DSC and AUC of the MICCAI 2019 Challenge on Digestive-System Pathological Detection and Segmentation. The results indicate that our approach outperforms other involved methods.}
    \label{tab:result-show}
\end{table*}

\subsection{Tissue Segmentation and Classification Comparisons}

On the unreleased test data of DigestPath2019, the final tissue segmentation and classification results are reported as Table \ref{tab:result-show}\footnote{The challenge result can be found in \url{http://www.digestpath-challenge.org/\#/}}. Our method achieves the best DSC and AUC (AUC is the same as the zju$\_$realdoctor team) at the same time, and we achieve the best \textbf{Final Rank} among all methods. This demonstrates that our model has strong generalization ability on the unknown test data. From Table \ref{tab:result-show}, we can observe that most of the methods achieve high AUC for WSI-level classification, but fail to obtain decent DSC for segmentation. This indicates that the lesion segmentation task is more challenging compared to WSI classification.

\begin{table}[!htbp]
    \centering
    \begin{tabular}{cccc}
\toprule
   $S$& LS & Patch Acc.& WSI AUC\\ \midrule
    10\% & -  &0.97526 &0.9990\\
    5\% & - &0.99357 & 0.993 \\
     {10\%}&  \checkmark &0.98538 &0.9981\\
     {5\%}&  \checkmark &0.99470 &1\\
\bottomrule
    \end{tabular}
    \caption{{WSI AUC and patch-level accuracy of proposed methods in Stage-1 on our validation dataset.}}
    \label{tab:stage1-1}
\end{table}

\subsection{WSI Classification}
Total 29504 patches are sampled to train the patch-level model used for WSI classification, and 3557 patched are utilized as the validation set. As denoted before, 68 WSIs are used to evaluate the WSI-level classification. Note that we label a patch as the positive sample when the malignant area ratio is bigger than a threshold $S$. In our experiment, we include two thresholds $S = 5\%$ and $S = 10\%$. We also validate the proposed label smoothing (LS) under $S = 10\%$. 

\begin{table}[!htbp]
    \centering
    \begin{tabular}{ccc}
\toprule
   LS & Patch Acc.& WSI AUC\\ \midrule
     -  &0.95234 &0.9330\\
       \checkmark &0.95618 &0.9450\\
\bottomrule
    \end{tabular}
    \caption{{WSI AUC and patch-level accuracy of proposed methods in Stage-1 on Camlyon16 dataset.}}
    \label{tab:stage1-1-cam}
\end{table}

{We summarize the WSI and patch classification results of Stage-1 in Table \ref{tab:stage1-1}. Note that the patch level accuracy is listed just for showing the performance of our model in the collected patch-level validation set. For patch level accuracy, our proposed method achieves 0.97526, 0.99357, 0.98538 and 0.99470 corresponding to: 1) LS close and $S = 10\%$; 2) LS close and $S = 5\%$; 3) LS open and $S = 10\%$; 4) LS open and $S = 5\%$, respectively. Smaller threshold $S$ can introduce higher patch accuracy under the same LS configuration, as denoted in Table \ref{tab:stage1-1}. Under the same threshold $S = 10\%$, the LS can bring 1\% accuracy increase, from 0.97526 to 0.98538, which validates the effectiveness of the proposed LS. Based on the patch level prediction results, the WSI classification is performed. Corresponding to the above four configurations, 0.9990, 0.993, 0.9981 and 1 WSI AUC are obtained on our validation set. We submitted the first three solutions (the fourth solution is conducted after the DigestPath2019 challenge) to the challenge, one of the solutions achieve WSI AUC 1 on the unknown testing dataset. Note that the observed testing phenomenon may be different between our defined validation set and the final testing set. These differences will not be discussed due to the unavailability of the testing set. We also tested our proposed LS scheme on Camlyon16 dataset, and listed the results in Table \ref{tab:stage1-1-cam}. On this dataset, our scheme can bring more than 1\% WSI AUC gain (from 0.9330 to 0.9450), which further verifies the effectiveness of our method. It should be noted that we follow the same image patch labeling in work \cite{li2018cancer}, where the ground truth label is determined by the center pixel label in the corresponding down-sampled patch, and thus $S$ is not applicable.}

When the positive WSIs are detected in Stage-1, we apply 3 patch models which trained by another sampled set to conduct key positive patch selection. The patch-level classification result in Stage-2 are listed in Table \ref{tab:stage2}, which denotes that our adopted multi-model voting scheme (Ensemble) obtains the best recall (0.9362), precision (0.9027) and accuracy (0.8935), respectively. This patch-level classification result is vital for the following segmentation. In our validation set, the recall reaches 0.9362 and the final segmentation result indicates this recall is sufficient.



\begin{table}[!htpb]
    \centering
    \begin{tabular}{cccc}
    \toprule
    Method & Recall &Precision & Accuracy\\ \midrule
    ResNet & 0.9293 &0.8804 &0.8721\\
    DenseNet& 0.9261&0.8986 & 0.8832\\
    ResNeXt& 0.9198 &0.8765& 0.8648\\
    Ensemble &\textbf{0.9362} &\textbf{0.9027}&\textbf{0.8935}\\
    \bottomrule
    \end{tabular}
    \caption{Recall, Precision and Accuracy of four methods: ResNet101, DenseNet161, ResNeXt101, and Ensemble. Ensemble means the voting result of adopted three models. Note that all the models are trained with patches cropped from positive WSIs.}
    \label{tab:stage2}
\end{table}

\subsection{Lesion Segmentation}
\begin{table}[!htpb]
\centering
\begin{tabular}{@{}ccc@{}}
\toprule
Method  & Patch            & WSI             \\ \midrule
Work \cite{mejbri2019deep}     & 0.8568            & 0.8120          \\
Work \cite{tao2019highly} & 0.8505                    & 0.7591          \\
Ours    & \textbf{0.8749}           & \textbf{0.8292} \\ \bottomrule
\end{tabular}
\caption{Segmentation performance comparisons among work \cite{mejbri2019deep}, work \cite{tao2019highly} and our proposed method.}
\label{tab:lesion-seg}
\end{table}
We compare our proposed method with another two WSI segmentation solutions: 1) directly performing segmentation on cropped patches based on UNet \cite{mejbri2019deep}; 2) performing patch classification first and then segmenting the selected positive patches \cite{tao2019highly}. Testing is presented on two validation datasets: 1) the sampled lesion patches from WSI; 2) the entire 68 validation WSIs. The lesion segmentation results are tabulated in Table \ref{tab:lesion-seg}. As presented in Table \ref{tab:lesion-seg}, our work achieves the best patch level and WSI level segmentation accuracy among all the listed methods. The reported accuracy of our work shows 1.74\% and 1.7\% accuracy gain than the second place schemes on patch and WSI datasets, respectively. We submit the proposed method to the challenge and achieve WSI DSC 0.8075 on the unknown testing dataset, which outperforms all the other methods. 

\begin{figure}[h]
\centering
\includegraphics[width=3.51in]{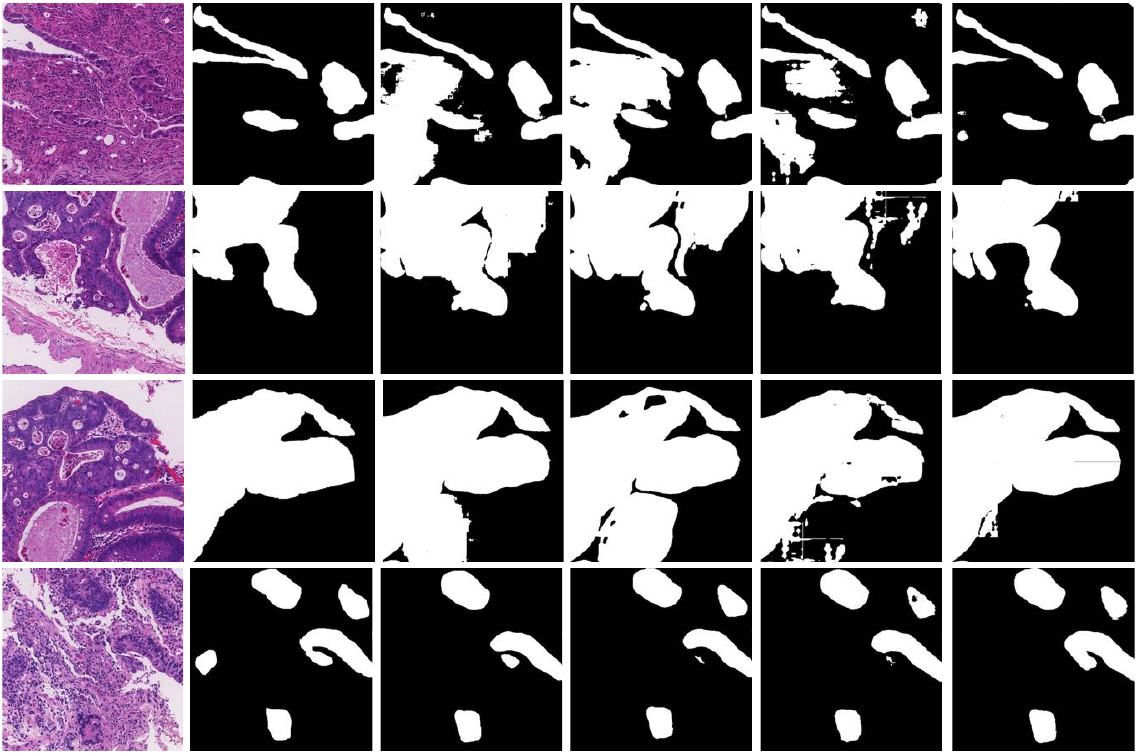}
\caption{Visual quality comparisons of image segmentation on four selected samples. From left column to right column: input images, ground truth, FCN, UNet, DeepLab, and our method.}
\label{fig:visual}
\end{figure} 

Some visual results of the segmented images by various algorithms are presented in Fig. \ref{fig:visual}. Obviously, our proposed method generates the best perceptual results. The proposed method not only can segment out each positive areas but also preserves much finer texture details, showing much better smooth and continuous visual boundaries than basic UNet, FCN and DeepLab \cite{chen2017deeplab}. 




\subsection{Ablation Study}
We further discuss the contributions of each component in our scheme to analyze the performance gain in detail. The ablation study includes two parts: the multi-level detection architecture and the segmentation techniques.
\begin{table}[!htpb]
\centering
\begin{tabular}{@{}cccc@{}}
\toprule
Method  &  1 Level    & 2 Levels       & 3 Levels             \\ \midrule
FPN     &  0.7058          & 0.7940          & 0.8097          \\
FCN     &  0.7550          & \textbf{0.7979} & 0.8177          \\
Deeplab &  \textbf{0.7591} & 0.7947          & 0.8134          \\
CAC-UNet    &  0.6663          & 0.7980          & \textbf{0.8292} \\ \bottomrule
\end{tabular}
\caption{Performance comparisons of four models (FPN, FCN, DeepLab, and CAC-UNet) under three architecture configurations: 1 Level, 2 Levels and 3 Levels.}
\label{tab:multi-level}
\end{table}

\textbf{Multi-level detection architecture.} In our proposed scheme, we adopt multi-level detection architecture: we first detect the positive WSIs (Stage-1), then detect the positive patches in positive WSIs (Stage-2), and finally segment the selected patches (Stage-3). Note in different stages, the proposed models and techniques are different. Four models are involved (FPN \cite{lin2017feature}, FCN, Deeplab, and CAC-UNet) for testing under 3 configurations: 1) directly segmenting all the patches of each WSI (1 Level); selecting the key patches and then segmenting the key patches (2 Levels); selecting key WSIs first, then choose key patches and finally segmenting the key patches (3 Levels). The results show that all the models with 3 Levels of architecture achieve the best performance when compared with the 1 Level and 2 Levels configurations, which indicates that the multi-level (three levels) detection architecture is very important for the segmentation of WSI. We believe that for a WSI with large resolutions, taking multi-level architecture having more advantages than the 1 Level or 2 Levels scheme. For example, we do not need to segment a negative WSI at all. However, if we directly crop this WSI into many patches, and then perform segmentation for these patches (or the selected patches when using the architecture of 2 Levels) one by one with segmenting model. This will bring risk for the detection of many false positive areas. We are convinced that the multi-level detection can conduct better results by flexibly applying different level information. Note that in Table \ref{tab:multi-level}, our proposed CAC-UNet performs poor in 1 Level and 2 Levels configuration. The reason is that our model is trained based on the positive patches selected from the positive WSIs, and this model may fail to detect the negative WSIs or the negative patches of the positive WSIs. To improve the performance under architectures of 1 Level or 2 Levels, the proposed CAC-UNet should be retrained using proper samples.

\begin{table*}[h]
\centering
\begin{tabular}{@{}lccccccc@{}}
\toprule
UNet                      & Aug.                       & IBN                       & Hypercolumn               & SCSE                      & PPM                       & Patch DSC  & WSI DSC   \\ \midrule
\checkmark &                           &                           &                           &                           &                           & 0.8490 & 0.8082 \\
\checkmark & \checkmark &                           &                           & \textbf{}                 &                           & 0.8568 & 0.8120 \\
\checkmark & \checkmark & \checkmark & \textbf{}                 &                           &                           & 0.8628 & 0.8158 \\
\checkmark & \checkmark & \checkmark & \checkmark &                           & \textbf{}                 & 0.8646 & 0.8162 \\
\checkmark & \checkmark & \checkmark & \checkmark & \checkmark &                           & 0.8706 & 0.8243 \\
\checkmark & \checkmark & \checkmark & \checkmark & \checkmark & \checkmark & 0.8726  & 0.8265      \\ \bottomrule
\end{tabular}
\caption{Performance gains by gradually integrating the adopted techniques (Aug., IBN, Hypercolumn, SCSE, and PPM) to UNet.}
\label{tab:seg-detail}
\end{table*}

\textbf{Segmentation Techniques.} In this ablation study, we demonstrate the effectiveness of different techniques used in our segmentation model. Table \ref{tab:seg-detail} shows performance gains when gradually adding the adopted techniques to UNet: data argumentation (Aug.), IBN, Hypercolumn, SCSE
, PPM. The backbone of our proposed CAC-UNet achieves 0.8726 patch DSC and 0.8265 WSI DSC. We further analyze the performance of different discriminators based on CAC-UNet, and list the related results in Table \ref{tab:dis}. The results denote that each discriminator can further boost the patch-level and WSI-level performance. When integrating   $D_e$, $D_d$ and $D_m$ together, the segmentation model produces the best performance, which indicates that these discriminators have some complementary nature in pathology image segmentation.

\begin{table}[h]
\centering
\begin{tabular}{@{}lcc@{}}
\toprule
Method & Patch DSC        & WSI DSC         \\ \midrule
$D_e$   & 0.8730       & 0.8275       \\
$D_d$   & 0.8731       & 0.8274       \\
$D_m$   & 0.8731       & 0.8275       \\
$D_e+D_d+D_m$ & {0.8749} & {0.8292} \\ \bottomrule
\end{tabular}
\caption{Performance comparisons by using different discriminators.}
\label{tab:dis}
\end{table}

\begin{table}[h]
\centering
\begin{tabular}{@{}lcc@{}}
\toprule
Method & Patch DSC        & WSI DSC         \\ \midrule
UNet   & 0.8513±0.0025       & 0.8096±0.0024       \\
Work \cite{tellez2018whole}   & 0.8521±0.0029       & 0.8102±0.0027       \\
Ours ($D_e+D_d+D_m$) & {0.8565±0.0034} & {0.8167±0.0031} \\ \bottomrule
\end{tabular}
\caption{{Performance comparisons between work \cite{tellez2018whole} and our domain adaptation scheme with three discriminators on DigestPath2019 dataset. Both work \cite{tellez2018whole} and our method are realized based on basic UNet.}}
\label{tab:dis-e1}
\end{table}

\begin{table}[h]
\centering
\begin{tabular}{@{}lc@{}}
\toprule
Method & Patch DSC                 \\ \midrule
UNet   & 0.9074±0.0025              \\
Work \cite{tsai2018learning}   & 0.9112±0.0026              \\
Work \cite{kamnitsas2017unsupervised} & {0.9159±0.0018}  \\ 
Our ($D_e$)   & 0.9174±0.0012              \\
Our ($D_d$)   & 0.9166±0.0015              \\
Our ($D_m$)   & 0.9153±0.0012              \\
Our ($D_e+D_d+D_m$)   & 0.9187±0.0013 \\
\bottomrule
\end{tabular}
\caption{{Performance comparisons with work \cite{tsai2018learning}, work \cite{kamnitsas2017unsupervised} and our domain adaptation scheme on renal biopsy pathology dataset. The PASM stained patches and the PAS stained patches are used as the source and target domains, respectively. To make fair comparisons, all involved schemes are reproduced (\cite{tsai2018learning} and \cite{kamnitsas2017unsupervised}) or realized (our scheme) based on the basic UNet.}}
\label{tab:dis-e2}
\end{table}

\begin{figure}[h]
\centering
\includegraphics[width=3.31in]{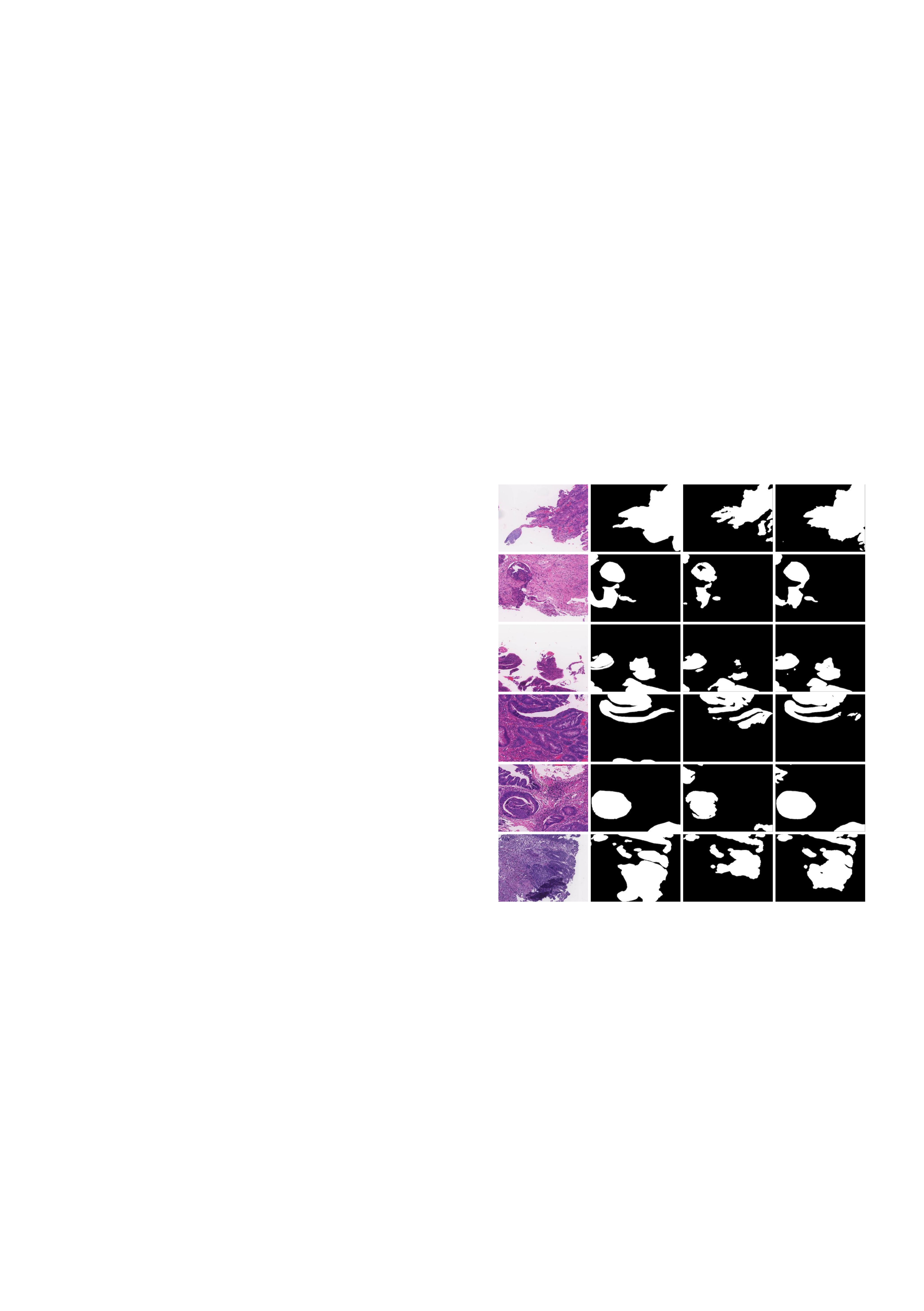}
\caption{{Performance verification of our proposed DA scheme on  DigestPath2019 dataset. Visual quality comparisons of image segmentation on six selected samples. From left column to right column: input images, ground truth, UNet, and UNet+($D_e$+$D_d$+$D_m$).}}
\label{fig:D1-vis}
\end{figure} 

\begin{figure}[h]
\centering
\includegraphics[width=3.31in]{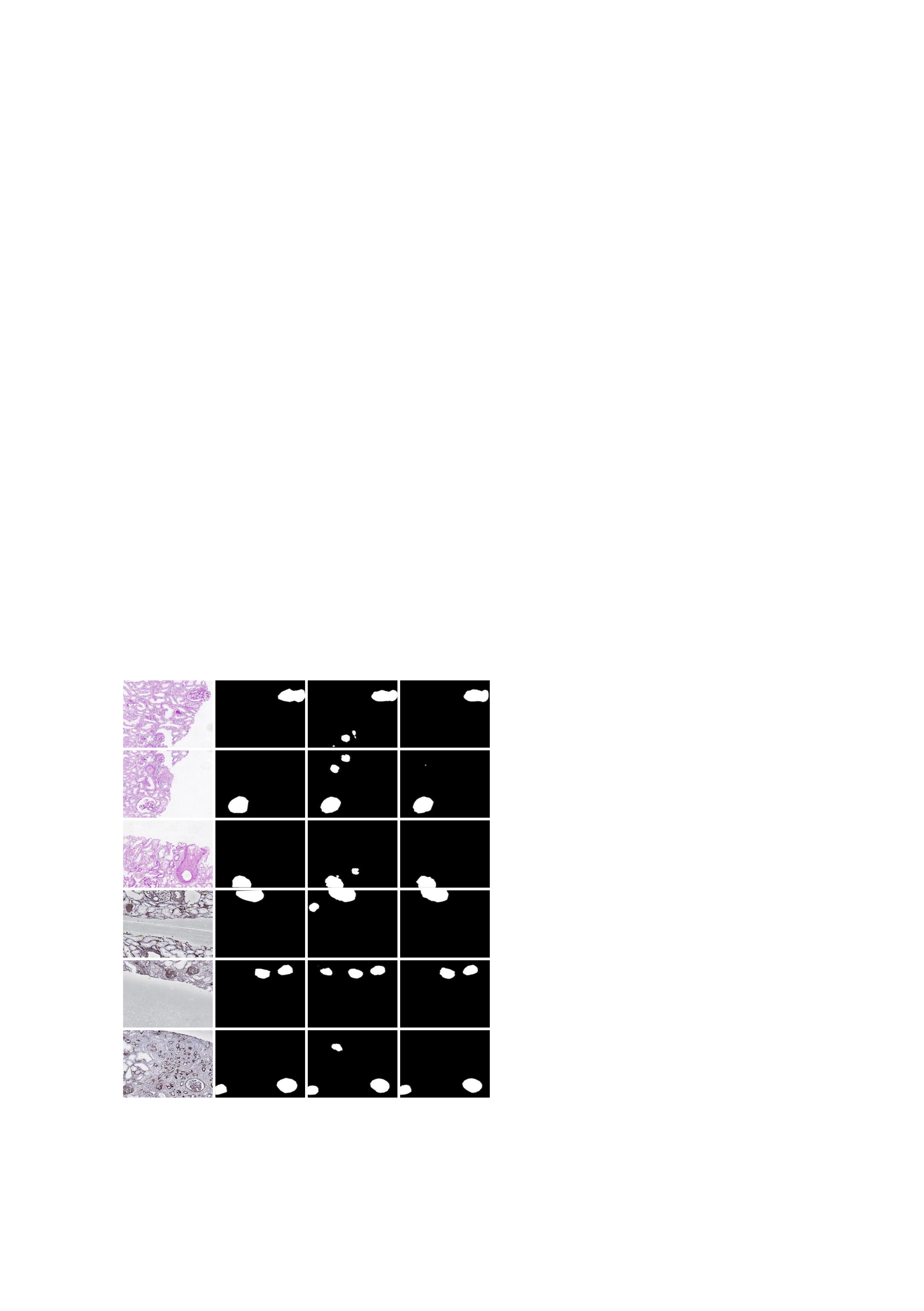}
\caption{{Performance verification of our proposed DA scheme on renal biopsy pathology dataset. Visual quality comparisons of image segmentation on six selected samples. From left column to right column: input images, ground truth, UNet, and UNet+($D_e$+$D_d$+$D_m$).}}
\label{fig:D2-vis}
\end{figure} 

{The above Table \ref{tab:dis} is our basic ablation experiment of adversarial learning scheme for image segmentation in DigestPath2019 challenge. To further verify the effectiveness of our method, we conduct two more experiments: (1) comparisons between our scheme and work \cite{tellez2018whole} which achieves domain invariance by using stain augmentation; (2) comparative analysis of our method and strategies in work \cite{tsai2018learning} and work \cite{kamnitsas2017unsupervised}. Different from the above comparisons, all results listed in Table \ref{tab:dis-e1} and Table \ref{tab:dis-e2} are the averages of 10 trials to more stably analyze the performance of our DA.}

{From Table \ref{tab:dis-e1}, we can see that our adversarial domain adaption method achieves better performance than work \cite{tellez2018whole} for both patch and WSI DSC. It should also be noted that our scheme obtains higher gains in UNet than in the proposed backbone of CAC-UNet (see Table \ref{tab:dis}). Part of the reason for this phenomenon is that our proposed backbone of CAC-UNet has already integrated some techniques that can learn appearance invariant features, such as the IBN block. On the renal biopsy pathology dataset, our scheme introduces about 0.0113 patch DSC gain (from 0.9074 to 0.9187) than the basic UNet, which shows superior results than the recent domain adaption work \cite{tsai2018learning} and work \cite{kamnitsas2017unsupervised}. We also verify our proposed DA scheme through visual quality comparisons on both DigestPath2019 and renal biopsy pathology dataset, as depicted by Fig. \ref{fig:D1-vis} and Fig. \ref{fig:D2-vis}. The visual quality comparisons also denote that our DA scheme can bring higher performance gain on the renal biopsy pathology dataset, which is similar to the objective performance in Table \ref{tab:dis-e1} and Table \ref{tab:dis-e2}.} 

\subsection{Computing Complexity Analysis}
\begin{figure}[tb]
\centering
\includegraphics[width=3.21in]{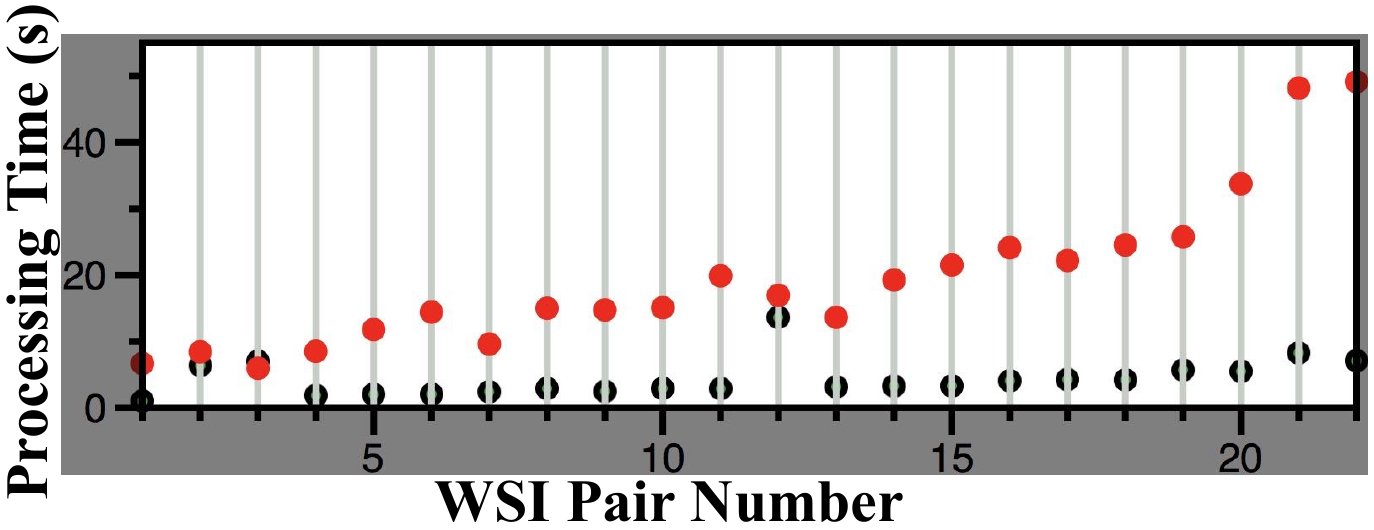}
\caption{Processing time of selected 22 WSI pairs (44 WSIs). The red and black points denote the positive WSIs and negative WSIs, respectively.}
\label{fig:time}
\end{figure}

On our 68 validation WSIs with size from 2371 $\times$ 1792 to 11246 $\times$ 23473, the average processing time is 15.3s, which is 
far below than the upper limit (120s) required by the challenge. In Fig. \ref{fig:time}, we depict the processing time for 22 pairs of WSIs. Each pair contains a positive WSI and a negative WSI, and the WSIs in the same pair have similar resolutions. We sort the WSI pairs in ascending order by their resolutions. As shown by Fig. \ref{fig:time}, with the increase of WSI resolutions, the processing time of positive WSIs (the red points) increases correspondingly with fast speed. However, the processing time of the negative WSIs (the black points) just rise slightly, and this is because most of the WSIs are dropped in Stage-1 and will not be processed in Stage-2 and Stage-3. Thus much processing time can be saved by the negative WSIs and then our scheme can focus on the processing of important positive WSIs. 

\section{Discussion}
\label{sec:discussion}
Automatic and objective medical diagnostic model can be valuable to achieve early cancer detection and diagnosis based on different pathological WSIs, and thus can reduce the mortality rate. The existing method directly apply the same patch-level model to perform both WSI-level and patch-level classification is hard to balance both tasks at the same time. Besides, the existing segmentation models are lack of the ability of appearance invariant. In this study, we proposed a highly efficient multi-level malignant tissue detection architecture, and designed an adversarial CAC-UNet to achieve robust appearance-invariant segmentation. 

\begin{figure}[tp]
\centering
\includegraphics[width=3.01in]{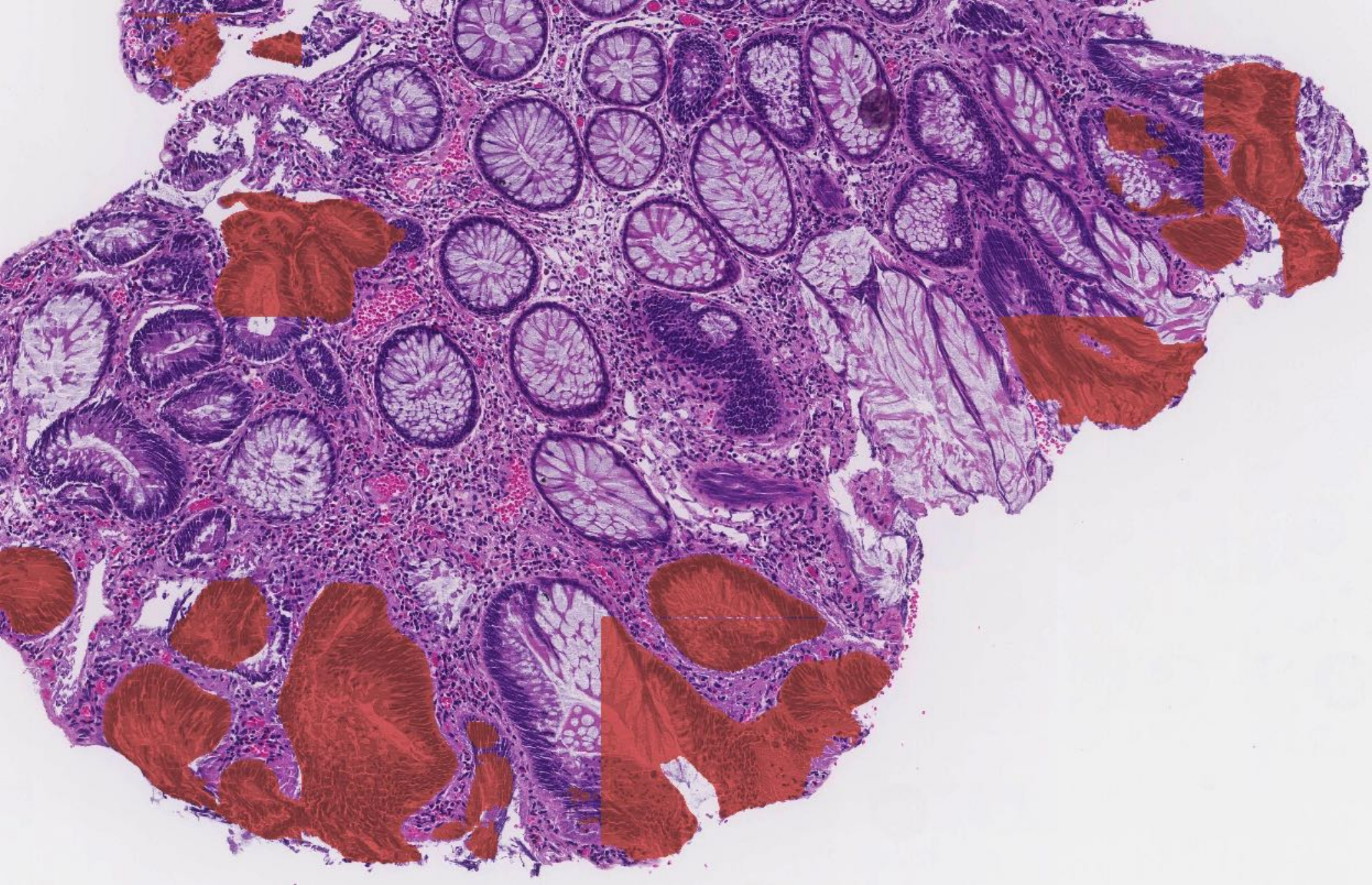}
\caption{False positive area illustration. Performing UNet on a cropped negative WSI, the red part denotes the detected false positive areas.}
\label{fig:risk}
\end{figure} 

We found that our proposed scheme achieves the best performance on DigestPath2019 colonoscopy tissue segmentation and classification task (see Table \ref{tab:result-show}), indicating the effectiveness of the proposed multi-level colonoscopy malignant tissue detection by using the designed adversarial CAC-UNet. Specifically, our proposed three-level detection can conduct better results than one-level and two-level architecture (see Table \ref{tab:multi-level}). This architecture can lower the risk of predicting many false positive areas. Fig. \ref{fig:risk} shows many false positive areas in a cropped negative WSI without any malignant tissue by using UNet, which confirms that the one-level architecture suffers false positive area detection. We also show that our proposed CRC-UNet backbone and the adversarial scheme can accurately conduct the tissue segmentation (see Table \ref{tab:lesion-seg}, Table \ref{tab:seg-detail} and Table \ref{tab:dis}).  

Our results provide compelling performance for colonoscopy malignant tissue detection through the proposed multi-level adversarial CAC-UNet. However, some limitations are worth noting. The key patch selection scheme of Stage-2 performs mediocrely on the validation dataset, which needs further improvement in the future. Besides, in this work we just divide the training set into two domains, however multiple domains should be studied for further increasing the generalization ability. Future work should therefore include to design stronger DA scheme. We release our codes in the GitHub to support the possible interested discussion.

\section{Conclusion}
\label{sec:conclusion}
We have presented a multi-level colonoscopy malignant tissue detection based on the proposed adversarial CAC-UNet, and we have shown that the proposed detection architecture with our segmentation model achieve superior results. The promising results and designed algorithms can be applied to automatic diagnosis scenario.

\section{Acknowledgement}
\label{sec:ack}
This work was supported in part by the Beijing Natural Science Foundation (4182044 and 7202056), the National Natural Science Foundation of China (81972248). This work is conducted on the platform of Center for Data Science of Beijing University of Posts and Telecommunications.


\bibliography{mybibfile}

\end{document}